\documentclass[aps,twocolumn,prd,showpacs,showpreprint,showkeys,nofootinbib,superscriptaddress,preprintnumbers]{revtex4}

\usepackage{graphicx,floatflt}
\usepackage{amsmath}
\usepackage{amssymb}

\newcommand{\dd}{\mathrm{d}} 
\newcommand{\Mpl}{M_\mathrm{pl}} 
\newcommand{\mpl}{m_\mathrm{pl}} 

\newcommand{\epsone}{\epsilon_1}
\newcommand{\epstwo}{\epsilon_2}
\newcommand{\rr}{\mathrm}

\newcommand{\ui}{\mathrm{i}}
\newcommand{\uhit}{\mathrm{hit}}
\newcommand{\uc}{\mathrm{c}}
\newcommand{\us}{\mathrm{s}}

\begin{document}

\preprint{ULB-TH/08-19}

\title{Avoiding the blue spectrum and the fine-tuning of initial conditions in hybrid inflation}

\author{S\'ebastien Clesse}
\email{seclesse@ulb.ac.be}
\affiliation{Service de Physique Th\'eorique, CP225, Universit\'e Libre de Bruxelles, \\
Bld du Triomphe, 1050 Brussels, Belgium}
\affiliation{Center for Particle Physics and Phenomenology, Louvain University,
2 chemin du cyclotron, 1348 Louvain-la-Neuve, Belgium}

\author{Jonathan Rocher}
\email{jrocher@ulb.ac.be}
\affiliation{Service de Physique Th\'eorique, CP225, Universit\'e Libre de Bruxelles, \\
Bld du Triomphe, 1050 Brussels, Belgium}

\date{\today}

\begin{abstract}
Hybrid inflation faces two well-known problems: the blue spectrum of
the non-supersymmetric version of the model and the fine-tuning of the
initial conditions of the fields leading to sufficient inflation to
account for the standard cosmological problems. They are
investigated by studying the exact two-fields dynamics instead of
assuming slow-roll.
When the field  values are restricted to be less than the
reduced Planck mass, a non-negligible part of the initial
condition space (around $15\%$ depending on potential parameters)
leads to successful inflation. Most of it is located outside the usual
inflationary valley and organized in continuous patterns instead of
being isolated as previously found. Their existence is explained and
their properties are studied. This shows that no excessive fine-tuning
is required for successful hybrid inflation. Moreover, by extending the
initial condition space to planckian-like or super-planckian values,
inflation becomes generically sufficiently long and can produce a
red-tilted scalar power spectrum due to slow-roll violations. The
robustness of these properties  is confirmed by conducting our analysis
on three other models of hybrid-type inflation in various framework:
``smooth'' and ``shifted'' inflation in SUSY and SUGRA, and ``radion
assisted'' gauge inflation. A high percentage of successful
inflation for smooth hybrid inflation (up to $80\%$) is observed.
\end{abstract}



\maketitle

\section{Introduction}
For almost a decade, the cosmic microwave background data have
supported a cosmological concordance model, in which
inflation~\cite{Starobinsky:1980te,Guth:1980zm,Linde:1981mu,Linde:2005ht,Inflation+25},
an early phase of accelerated expansion, is the favored
explanation for the origin of the primordial
fluctuations\footnote{Note however that some alternatives exist
such as string gas cosmology~\cite{Brandenberger:2006vv} or
bouncing universe models \cite{Peter:2006hx,Peter:2008qz}.}. For
more than 25 years, many models of inflation have been proposed,
from toy models to more realistic models based on various high
energy physics frameworks
\cite{Lyth:1998xn,Linde:2005ht,Linde:2005dd,Kallosh:2007ig}. The
incoming flow of cosmological data has however started to
discriminate among the models. In particular, the last release of
the WMAP 5-years data~\cite{Komatsu:2008hk} favored a red tilted
scalar power spectrum.

If some single-field models are still able to reproduce the
current data, the presence of multiple scalar fields in all the
high energy physics frameworks proposed today (Higgs fields in
Grand Unified Theories, super-partners in Supersymmetry, moduli in
string theory) makes it hard to imagine that the inflaton field is
not coupled to any other scalars. The simplest (and yet motivated)
known example of multi-field inflation is the hybrid one. The
original hybrid model of inflation, proposed
in~\cite{Linde:1993cn,Copeland:1994vg}, had been introduced as an
alternative way to end inflation and could be realized for sub-planckian
field values unlike chaotic models. The key idea is to couple the inflaton field to a
Higgs-type waterfall field which ends inflation by acquiring a
non-vanishing Vacuum Expectation Value (vev). This model could be
considered as realistic if employing a Higgs field and an extra
singlet of some minimal extension of the Standard Model of
particle physics. It also represents a toy model for many
multi-field models of inflation in other frameworks. Indeed,
hybrid(-type) models of inflation have been embedded in almost all
high energy frameworks: in (extended) supersymmetry and
supergravity~\cite{Halyo:1996pp,Binetruy:1996xj,Dvali:1994ms,
Kallosh:2003ux}, in Grand Unified Theories~\cite{Jeannerot:1997is,
Jeannerot:2003qv}, or various extra-dimensional
theories~\cite{Dvali:1998pa,Koyama:2003yc,Fukuyama:2008dv,
Fairbairn:2003yx}.

When confronting the original hybrid inflation to the CMB data, it
is however well known~\cite{Lyth:1998xn} that the power spectrum
tilt is blue which is now disfavored. This is only valid
when slow-roll is assumed and when the vacuum energy density
dominates the potential since in the other case, the potential
becomes equivalent to a chaotic model. In this paper, we
illustrate these properties and study the predictions for the
spectral index of the model using the exact field dynamics.
We find a new way to generate a red tilted spectrum due to
non-trivial effects of the violation of slow-roll,
give the two possible conditions on the parameters of the
potential to generate a red spectrum of perturbations and discuss
the field values that these conditions require.\\

Several fundamental questions about initial conditions for
inflation are still open (see for
example~\cite{Goldwirth:1991rj,Vachaspati:1998dy,Kaloper:2002cs,
Linde:1986fd,Tetradis:1997kp,Lazarides:1997vv,Mendes:2000sq}).
In this paper, we will not address the important problem of spatial
homogeneity of the fields~\cite{Goldwirth:1991rj}
and we will assume that the field values do not enter the self-reproducing
inflationary regime~\cite{Linde:1986fd}. Even when restricting to
the classical approximation, the existence of a
fine-tuning on the initial values of the fields was found, for
hybrid
inflation~\cite{Tetradis:1997kp,Lazarides:1997vv,Mendes:2000sq}.
(An opposite conclusion has been obtained~\cite{Lazarides:1996rk}
for the smooth hybrid inflation model. We will comment on this
model at section~\ref{sec:othermodels}). The space of initial
conditions is described by regions in the plane
$(\phi_\ui,\psi_\ui)$, where $\phi_\ui$ and $\psi_\ui$ denote the
initial values of the inflaton and the waterfall field
respectively. By fine-tuning of the initial conditions, one means
that the regions leading to sufficient (60 e-folds or more)
inflation have been found to be
composed~\cite{Tetradis:1997kp,Mendes:2000sq} of an extremely thin
band around $\psi_\ui=0$ and a few apparently random points in the
rest of the plane. Uncertainties remain on whether these points are
of null measure~\cite{Mendes:2000sq} or not~\cite{Tetradis:1997kp}.
The thin band is also considered as fine-tuned
because $\psi_\ui$ has to be so close to 0 that any
quantum fluctuations would shift its value outside the successful
region~\cite{Tetradis:1997kp}. This would be an important problem
for hybrid-type inflation because it means that these models would
not easily be the natural outcome of some pre-inflationary era
(see however~\cite{Calzetta:1992bp}).

Several papers have proposed some solutions to the fine-tuning
problems. It has been proposed to replicate many times identically
the inflaton sector~\cite{Mendes:2000sq}, even though no
motivations have been proposed for this replication. A similar
idea had been employed to construct the N-flation
model~\cite{Dimopoulos:2005ac} but the replication in this context
is not more natural~\cite{Easther:2005zr}. It has also been
proposed~\cite{Mendes:2000sq,Underwood:2008dh} to embed hybrid
inflation into a brane description. The induced modifications to
the Friedmann-Lema\^itre equations provide additional friction in
the evolution of scalar fields. Thus slow-rolling is favored and
more of the initial condition space gives rise to successful
inflation. This friction can also be efficiently played by
dissipative effects~\cite{Ramos:2001zw}, when couplings between
the inflaton and the waterfall field with a bath of other fields
are assumed. Finally, it has been
proposed~\cite{Panagiotakopoulos:1997if} to solve this problem by
accepting a short ($N\sim 10$) phase of hybrid inflation and
implementing a second one responsible for the generation of the
primordial fluctuations, thus solving the horizon problem.

However, to our knowledge, little has been proposed to explain the
properties of the (un-)successful space of initial conditions:
discreteness, sub-dominance, size and limits. In this paper, we first
show that super-planckian initial conditions always give rise to a
sufficiently long phase of hybrid inflation and can produce a
red-tilted power spectrum without the need of any fine-tuning. We
provide a detailed analysis of the properties of the initial condition
space, explain why parts of this space were thought to be
discrete, and what are the field trajectories leading to these
apparently isolated points. In particular, we show that they can be
viewed as the ``anamorphosis'' (that is a deformed image) of the
thin successful band. We also give the area of successful initial
conditions  in the the plane $(\phi_\ui,\psi_\ui)$. Even when
restricting the fields to sub-planckian values, we find around $15\%$
of successful initial conditions and we discuss the effect of varying
the different parameters of the potential. When going to
super-planckian values, we confirm that this ratio tends to $100\%$.

To prove the robustness of these results, we explore the space of
initial conditions for three other hybrid-type models: the
supersymmetric and supergravity
``smooth''~\cite{Lazarides:1995vr,Lazarides:2007fh,Yamaguchi:2004tn}
and ``shifted''~\cite{Jeannerot:2000sv,Jeannerot:2002wt} models,
as well as the ``radion assisted'' gauge
inflation~\cite{Fairbairn:2003yx}. The first two models are direct
extensions of the F-term hybrid inflation \cite{Dvali:1994ms} and
are motivated by the fact that their inflationary
valley is shifted away from $\psi=0$, so that any harmful
topological defects formed during the symmetry breaking induced by
$\psi$ would be diluted away. The last model is based on a
hybrid-type potential even though constructed in 5D. Its main
motivation resides in the fact, that by construction, the form of
the potential is controlled (and thus protected) by gauge
symmetries.

Before going any further, let us discuss the physical motivations
of enlarging the space of initial conditions to super-planckian
values. Even though this possibility has been moderately studied
in \cite{Lazarides:1997vv}, most previous
works~\cite{Tetradis:1997kp,Mendes:2000sq} restricted their
analysis to initial values of the fields under the Planck mass.
For non-supersymmetric four-dimensional theories, it was proposed
in~\cite{Linde:2005ht} that quantum gravity corrections are
controlled as long as the energy density and the effective masses
are sub-planckian. In this case, effective field theories can be
an appropriate framework to describe fields of planckian-like
amplitude.
More recently, several models such as natural
inflation~\cite{Freese:1990rb}, or gauge
inflation~\cite{ArkaniHamed:2003wu,ArkaniHamed:2003mz,Kaplan:2003aj}
also allow fields to be super-plankian. For example, in gauge
inflation, the inflaton field is part of a gauge field and thus
the form of the potential is protected by gauge symmetries,
leaving non-renormalizable corrections highly constrained.

Let's turn to supersymmetric frameworks. Since uncontrolled
non-renormalizable corrections to the superpotential and the
K\"ahler potential appear, super-planckian fields are inevitably
problematic for models constructed in the context of supersymmetry
(SUSY) or supergravity (SUGRA). Global SUSY is only valid as long
as all fields have an amplitude much smaller
than\footnote{Throughout this paper, we will denote the Planck
mass by $\mpl\equiv G^{-1/2}\simeq 1.2 \times 10^{19}\;
\mathrm{GeV}$ and the reduced Planck mass by $\Mpl\equiv (8\pi
G)^{-1/2}$.} $\Mpl$. When closer to the reduced Planck mass (but
still below), SUGRA corrections are important and supergravity is
the correct framework to describe the model. Above $\Mpl$, the
non-renormalizable corrections become dominant: the
non-renormalizability of  SUGRA prevents us from using it and a
UV-complete theory is then necessary \cite{Baumann:2009ni}.
Finding and describing
inflaton fields with Planckian displacements in string theories is,
however, possible in certain sectors of string theories, though not always
stable and their potential is not easily flat \cite{Baumann:2009ni,Hsu:2003cy}.
Several models of inflation constructed within string theories have
been proposed (see for e.g.
\cite{HenryTye:2006uv,Kallosh:2007ig,Silverstein:2008sg,Baumann:2009ni}
and refs therein) and some of them have a low-energy effective
description that mimics
hybrid inflation~\cite{Dvali:1998pa,Linde:2005dd,Koyama:2003yc}.
\emph{These examples motivated us to assume that the effective
inflationary potentials studied in this paper can also originate from
frameworks in which it is safe to consider super-planckian fields.}
As a conclusion, we have chosen to study them both in the sub-planckian
and super-planckian field regimes. However, we will always restrict
ourselves to sub-planckian energy densities.

The rest of the paper is organized as follows. In the
section~\ref{sec:originalhybrid}, we extensively study the
original hybrid inflation model~\cite{Linde:1993cn}. In
particular, we discuss the validity of the one-field slow-roll
approximation and show that violation of slow-roll conditions can
strongly modify the dynamics. Using exact numerical methods,
applied on the two-field potential, we provide a complete analysis
of the space of initial conditions and revisit the above-mentioned
fine-tuning problem. In Sec.~\ref{sec:othermodels}, we test the
robustness of our results on the three other models: SUSY/SUGRA smooth
hybrid inflation, SUSY/SUGRA shifted hybrid inflation and radion-inflation.
Our conclusions are drawn in Sec.~\ref{sec:conclu} and some open
questions are developed.

\section{Original hybrid model}\label{sec:originalhybrid}
Proposed in~\cite{Linde:1993cn,Copeland:1994vg}, the model is
based on the potential
\begin{equation} \label{eq:potenhyb2d}
V(\phi,\psi) = \frac 1 2 m^2 \phi^2 + \frac \lambda 4 \left(\psi^2
- M^2 \right)^2 +\frac{\lambda'}{2} \phi^2 \psi^2,
\end{equation}
where $\phi$ is the inflaton and $\psi$ is the Higgs-type field.
$\lambda$ and $\lambda'$ are two positive coupling constants, $m$
and $M$ are two mass parameters.  It is the most general form
(omitting a quartic term $\lambda'' \phi^4$) of renormalizable
potential verifying the symmetries $\psi \leftrightarrow -\psi$
and $\phi \leftrightarrow -\phi $. In the general case, inflation
is mostly assumed to be realized in the false-vacuum along the
$\psi=0$ valley and ends with a tachyonic instability for the
Higgs-type field. The critical point of instability below which
the potential develops non-vanishing minima is
\begin{equation}
\phi_{\rr c} = M \sqrt{\frac {\lambda}{\lambda'}}\,.
\end{equation}
The system then evolves toward its true minimum at $V=0$,
$\langle\phi\rangle=0$, and $\langle\psi\rangle=\pm M$, where
throughout the paper, $\langle.\rangle$ denotes the vacuum expectation
value (vev) of a field.

In this section, we will first restrict ourselves to the effective
one-field approach to re-analyze the predictions of the spectral index
for the generated power spectrum. This will be done solving
numerically the exact field equations of motion rather than assuming
slow-roll. Then, we will  move one the full two-field dynamics
and study the initial conditions that lead to sufficiently long
inflation.

\subsection{Effective one-field potential}
To study the inflationary phase along the valley $\psi=0$, it is
common usage to restrict the potential of
Eq.~(\ref{eq:potenhyb2d}) to a one-dimensional effective potential
of the form
\begin{equation} \label{eq:potenhybeffectif}
V(\phi) =  \Lambda^4 \left[1+ \left( \frac{\phi}{\mu}
\right)^2\right],
\end{equation}
with
\begin{equation}\label{eq:definmu}
\mu \equiv \sqrt{\frac{\lambda}{2}} \frac{M^2}{m}~, \quad
\Lambda \equiv \lambda ^{1/4} M /\sqrt{2}~.
\end{equation}
The Friedmann-Lema\^itre equations and the Klein-Gordon equation in
an expanding universe for this scalar field read
\begin{equation} \label{eq:FLtc1}
\begin{split}
H^2 &= \frac {8\pi }{3 \mpl ^2 }  \left[ \frac 1 2 \dot \phi^2  +
V(\phi) \right], \\
\frac {\ddot a }{a} &= \frac {8\pi}{3 \mpl^2} \left[ - \dot
\phi^2+ V(\phi) \right],
\end{split}
\end{equation}
\begin{equation} \label{eq:KGtc}
\ddot \phi + 3 H \dot \phi + \frac {\dd V}{\dd \phi} = 0,
\end{equation}
where $a$ is the scale factor, $H$ the Hubble parameter, and the
dot denotes a derivative with respect to cosmic time. Inflation is
realized starting at high field values, the field rolling down
toward 0. To mimic the two-field dynamics, in this effective
model, it is necessary to define an effective critical value
$\phi_\uc >0$ at which inflation ends.

The well-known slow-roll approximation consists in neglecting the
second derivative of the field in Eq.~(\ref{eq:KGtc}) and the kinetic
terms compared to the potential in Eq.~(\ref{eq:FLtc1}).   Using
 the Hubble flow parameters~\cite{Schwarz:2001vv,Schwarz:2004tz}
\begin{equation}
\epsilon_0 \equiv H,
\quad  \epsilon_{\rr {n+1}} \equiv  \frac {\dd
\ln \epsilon_{\rr n}}{\dd N},
\end{equation}
inflation occurs for $\epsone=-\dot{H}/H^2 < 1$ and slow-roll
conditions are satisfied when, for all $n\geq 1$, $|\epsilon_n| \ll
1$. For the effective hybrid potential, an analytical expression of
the first and second Hubble flow parameter is easily derived in the
slow-roll approximation,
\begin{equation}\label{eq:epshybrid}
\begin{split}
\epsone (\phi) & =\frac{1}{4 \pi} \left( \frac {\mpl}{\mu} \right)^2
\frac{(\phi/\mu)^{2}}{\left[ 1 + (\phi/\mu)^2 \right]^2}\,, \\
\epstwo(\phi)&=\frac{1}{2\pi}\left(\frac{\mpl}{\mu}\right)^2
\frac{(\phi/\mu)^2 -1}{\left[1+(\phi/\mu)^2\right]^2}\,.
\end{split}
\end{equation}
It is clear from these expressions that the slow-roll conditions
are satisfied for large field values. As illustrated in
Fig.~\ref{fig:eps1_srviolated},
Eq.~(\ref{eq:epshybrid}) suggests that two phases of inflation can
take place~\cite{Martin:2006rs}. A first phase at large values of
the field, and a second phase at small values.  These phases are
separated by a maximum of $\epsone(\phi)$, reached at
$\phi_\mathrm{max}$, and at which $\epstwo(\phi)$ changes its
sign. In the slow-roll approximation $\phi_\mathrm{max}=\mu$.
However, around $\phi_\mathrm{max}$, and for sufficiently small
values of $\mu$, slow-roll conditions can be violated, as it is
illustrated by the dashed line in
Fig.~\ref{fig:eps1_srviolated}. Thus a resolution of the exact
equations of motion for the fields is required to study the
influence of the transition period on the dynamics of inflation.

\subsubsection{Exact field dynamics}\label{sec:1fielddynamics}

\begin{center}
\begin{figure}[ht]
\scalebox{.83}{\includegraphics{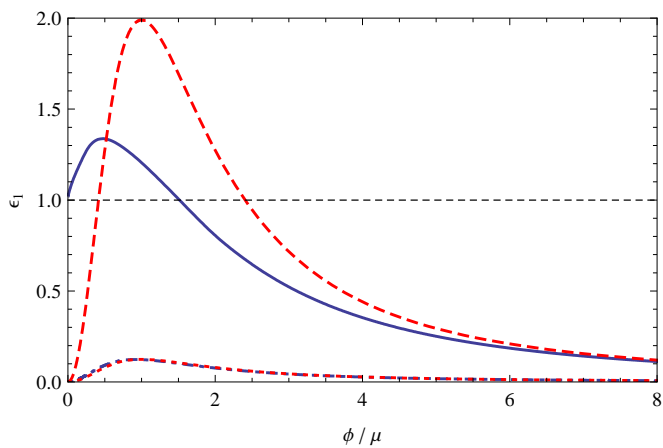}}
\caption{First Hubble-flow parameter $\epsone$, function of the
inflaton field, during its evolution started in the large field
phase, in the slow-roll approximation (red dashed and dotted
lines) and from the exact dynamics (blue solid and dot-dashed
lines).  The curves correspond to $\mu=0.1
\mpl$ (two top curves), $\mu=0.4 \mpl$ (two bottom curves,
quasi-superimposed). For each value of $\mu$, we observe two
phases of evolution at large field and at small field (compared
to $\phi_{\rm max}$).} \label{fig:eps1_srviolated}
\end{figure}
\end{center}

The dynamics of the one-field effective hybrid inflation, without
assuming slow-roll, is described by Eqs.~(\ref{eq:FLtc1}) and
(\ref{eq:KGtc}):they have been integrated numerically. The
parameter $\epsone$ has been computed exactly and
is represented as a function of the inflaton field in
Fig.~\ref{fig:eps1_srviolated}  and compared to the analytical
slow-roll expressions of Eq.~(\ref{eq:epshybrid}).

The exact integration confirms the existence of the two regimes
before and after the maximum of $\epsone$, at which the slow-roll
conditions can be violated and inflation can even be interrupted
(when $\epsone \geq 1$)
depending on the value of the parameter $\mu$. But there are two
important novelties. Firstly, $\phi_\mathrm{max}$ is displaced
toward \emph{smaller} values in the exact treatment compared to
its slow-roll value $\mu$. Secondly, in the slow-roll
approximation, after the peak, $\epsone(\phi)$ decreases and
vanishes for vanishing field. One may think that inflation always
takes place for $\phi< \phi_\rr{max}$. However, exact numerical
results show that this conclusion is erroneous: $\epsone$ does not
necessarily become negligible when the field vanishes (see the
plain blue curve). As a consequence, \emph{inflation does not
necessarily produce the last $60$ e-folds in the small field
regime ($\phi<\phi_{\max}$)}. \\

\begin{figure}
\scalebox{.3}{\includegraphics{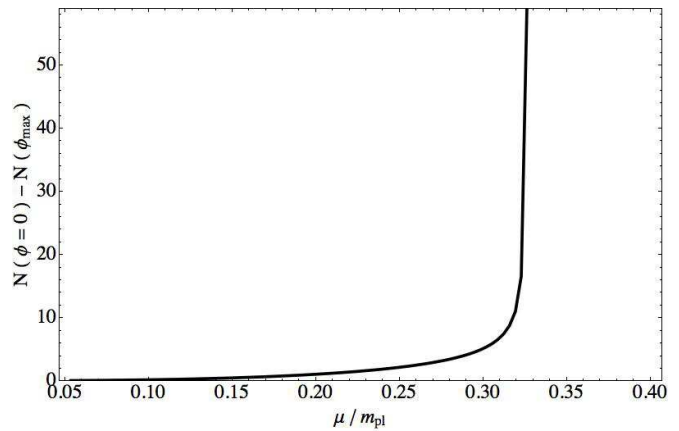}}
\caption{Number of e-folds created between $\phi_\rr{max}$ and
$\phi=0$ as a function of $\mu$, when slow-roll is not assumed.
There exist a critical value of the parameter $\mu$ under which
a marginal number of e-folds is generated in the second phase
of evolution.
Above the critical value of $\mu$, the number of e-folds created
in the second phase of inflation diverges showing the efficiency
of the second phase of inflation when it exists.}
\label{fig:criticalmu}
\end{figure}

From Fig.~\ref{fig:eps1_srviolated}, it is clear that the presence
or not of small field phase of inflation depend on the parameter
$\mu$ (difference between the dashed and plain curves). In order
to measure the efficiency/existence of this second phase of
inflation, we have plotted in Fig.~\ref{fig:criticalmu} the number
of e-folds created between $\phi_\rr{max}$ and $\phi=0$ as a
function of $\mu$.
This shows that there exists a critical value
\begin{equation}\label{cond_redspectrum_2}
\mu_{\mathrm{crit}}\simeq 0.32 \,\mpl,
\end{equation}
under which the number of e-folds generated after $\phi_\rr{max}$
is reached is marginal. In this case, the period of inflation
where the observable modes become super-Hubble will always take
place in the large field phase  ($\phi>\phi_{\max}$) provided
$\phi_\ui > \phi_{\max}$. In this case, the potential of hybrid
inflation leads to a chaotic-like inflation, independently of the
way inflation ends. This has important consequences for the
generated spectral index.

\subsubsection{Scalar spectral index}
At first order in slow-roll parameters, the spectral index of the
scalar power spectrum $\mathcal{P}_\zeta$ can be expressed
as~\cite{Schwarz:2001vv,Martin:2003bt}
\begin{equation}
n_\mathrm{s} -1 \equiv \left.\frac{\dd
\mathcal{P}_\zeta}{\dd \ln k}\right|_{k=k_*} = - 2 \epsilon_{1*} -
\epsilon_{2*}~.
\end{equation}
A star means that the quantity is evaluated at Hubble
crossing $aH=k_*$, $k_*$ being a pivot scale in the range of
observable modes.

Recent experimental results from WMAP 5-years~\cite{Komatsu:2008hk}
have a best fit at $n_\us\simeq 0.96$ and disfavor a value of the
scalar spectral index greater than unity at almost $95\%$ confidence
level (CL). From this observation, hybrid models have recently been
considered as disfavored. Indeed, in the slow-rolling effective
one-field model, as shown at the previous section, the last $60$ e-folds
of inflation are realized in the small field phase characterized by a
negative $\epstwo$ and a negligibly small $\epsone$ which induces
necessarily a blue spectrum.

\begin{figure}
\scalebox{.18}{\includegraphics{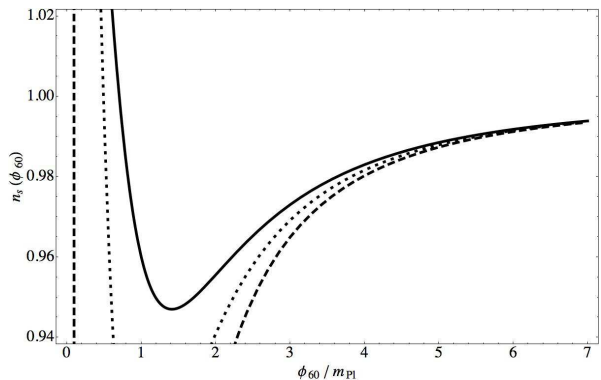}}\caption{Spectral
index $n_\us$ of the power spectrum as a function of $\phi_{60}$, the
value of the field $60$ e-folds before the end of inflation.  This has
been computed for the effective hybrid potential for $\mu=\mpl$ (full
line), $\mu=0.7 \mpl$ (dotted line) and $\mu = 0.14 \mpl$ (dashed
line), in the slow-roll approximation. One can see that almost any
value of the spectral index can be accommodated  within hybrid
inflation.} \label{fig:ns}
\end{figure}
However, there
exist two mechanisms to produce a red spectrum within the standard
hybrid inflation model along the valley $\psi=0$. There are two
ways of forcing the small field inflation phase not to take place,
either by instability or by violation of the slow-roll condition.
In both case, the consequence is that the spectral index generated
is below 1 as represented in Fig.~\ref{fig:ns}. Note that almost
any values of the spectral index can be actually
accommodated by the model, including the best fit for WMAP5 data.

\paragraph{When the critical point of instability is in the large field
phase.} The simplest way to obtain a red spectrum is to destabilize
inflation with the waterfall field at some stage during the large
field phase or at most at the peak, $\phi_\uc\geq\phi_{\rr{max}}$,
independently of $\mu$. This had been noticed in the past
\cite{Copeland:1994vg,Cardoso:2006wf}, though not often emphasized.
For $\phi \gg \phi_{\rr {max}}$,
the inflaton potential is of the form $V\simeq m^2 \phi^2/2$ for
which $n_\us<1$. Since in the exact treatment $\phi_{\rr {max}} <
\mu$ is shifted to smaller values, a sufficient condition to have
$n_\us<1$ reads
\begin{equation}\label{cond_redspectrum_1}
\frac{m}{M} > \sqrt{\frac{\lambda'}{2}}~.
\end{equation}
This property still holds when violations of slow-roll are taken into
account.

We would like to emphasize that the value of the inflaton $60$ e-folds
before the end of inflation, denoted $\phi_{60}$, is necessarily
super-planckian if inflation takes place in the large field regime,
independently of $\mu$.  If $\mu \geq \mu_{\rm crit}$, the slow-roll
approximations can be used and with $\phi_{\rm end}=
\phi_{\uc}=\phi_{\rr {max}}=\mu$, it is well
known~\cite{Copeland:1994vg} that the minimum
value of $\phi_{60}$ is given by
\begin{equation}
\frac{2\pi\mu^2}{\mpl^2}\left[2\ln\left(\frac{\phi_{60}}{\mu}\right)
+ \left(\frac{\phi_{60}}{\mu}\right)^2-1\right]=N_{\rr{succ}}=60,
\end{equation}
which is always around $3\mpl$ or greater. If $\mu \leq \mu_{\rm crit}$,
solving numerically the exact field dynamics is required, and we also
found that $\phi_{60}\gtrsim 3\mpl$.

\paragraph{When the second phase never takes place.}
Assuming that the initial value is in the large field phase
$\phi_i > \phi_\mathrm{max}$, if $\mu\lesssim\mu_{\rm crit}$,
then the small field phase of inflation can never take place.
This is a new way to generate a red spectrum independently of
the critical value $\phi_{\rm c}$. Indeed,
an excessive velocity of the field around $\phi_\mathrm{max}$
induces a violation of the (slow-roll) inflation condition (see
plain line of Fig.~\ref{fig:eps1_srviolated}). Thus
$\phi_{60}$ necessarily lies in the large field regime and the
spectrum is red, independently of the critical value $\phi_\uc$.
Notice however that even for $\mu \le \mu_{\mathrm{crit}}$, one
could still start an inflationary period with $\phi_\ui \ll
\phi_{\max}$ leading to a blue-tilted power spectrum. Violation of
slow-roll only prevents this period to occurs after any large
field phase. In this case also, this requires a large initial
value of the inflaton, and a realization of hybrid inflation in a
regime away from the usual limit $\phi \ll \mu$.  This conclusion might
reduce the appeal of the model.

\subsection{Exact two-field dynamics and initial conditions}
We now turn to the two-field potential given in
Eq.~(\ref{eq:potenhyb2d}) to study the field dynamics without
restricting to the $\psi=0$ valley. In previous works,
Tetradis~\cite{Tetradis:1997kp}, Lazarides \&
Vlachos~\cite{Lazarides:1997vv} and more recently Mendes \&
Liddle~\cite{Mendes:2000sq} studied the space of initial
conditions of the fields leading to successful/unsuccessful
inflation for hybrid inflation. They found that the successful
regions for sub-planckian initial values are made of a very narrow
band along the $\psi = 0$ axis (motivating the one-field
approach), together with some scattered points in the unsuccessful
region, which seemed randomly distributed. In this section, we
explore a larger space of initial conditions and extend previous
studies to super-planckian initial values. We show that three
different classes of successful trajectories in field space can be
defined, one of them explaining the origin and the properties of
the previously found isolated points. Finally, we quantify the
amount of fine-tuning of the model by computing the ratio of
successful/unsuccessful area and study the effect of varying the
parameters of the potential on our results.

\subsubsection{Exact two-field dynamics}
For two homogeneous scalar fields $\phi$ and $\psi$, the
Friedmann-Lema\^itre equations take the form
\begin{equation} \label{eq:FLtc12field}
\begin{split}
H^2 &= \frac {8\pi }{3 \mpl^2}  \left[ \frac 1 2 \left(\dot
\phi^2 + \dot \psi^2 \right)  + V(\phi,\psi) \right], \\
\frac{\ddot a }{a} &= \frac {8\pi}{3 \mpl^2} \left[ - \dot \phi^2
- \dot \psi^2 + V(\phi,\psi ) \right],
\end{split}
\end{equation}
while the equations of Klein-Gordon for these scalar fields read
\begin{equation} \label{eq:KGtc2field}
\begin{split}
&\ddot \phi + 3 H \dot \phi + \frac {\partial
V(\phi,\psi)}{\partial \phi} = 0, \\
&\ddot \psi + 3 H \dot \psi + \frac {\partial
V(\phi,\psi)}{\partial \psi} = 0.
\end{split}
\end{equation}
For the numerical integration, instead of using the scale factor
and its time derivative as integration variables, it is more
convenient to use the number of e-folds realized from the
beginning of inflation\footnote{$a_{\rr i}$ is the scale factor at
the beginning of inflation} $N(t)=\ln\left[ a(t)/a_{\rr i}
\right]$ and its first derivative - the Hubble parameter.

\subsubsection{Classical dynamics and stochastic effects}
Considering large values for the fields can induce stochastic
(quantum) effects to affect the field dynamics, described in this
paper as purely classical~\cite{Linde:1993xx,Martin:2005ir}.
Since we also consider super-planckian field values, it is important to
check that for such values, the dynamics is still dominated by the
classical motion. The stochastic effects in the full two-field
potential have not yet been studied but
the stochastic effects should be very limited. Indeed, the
dynamics that is found is fast-rolling at the beginning, during
which the classical motion will clearly dominate and then
slow-roll in the inflationary valley. When slow-roll is realized,
it is possible to evaluate at what field values the stochastic
effects become relevant by comparing the classical field
fluctuations and the quantum field fluctuations, during a Hubble
time. In the valley $\psi=0$, we obtain
\begin{equation}
\frac{H}{2\pi}\gtrsim \frac{\Mpl^2 V'}{V} \quad \Leftrightarrow
\quad \frac{\phi}{\mpl^2} \gtrsim \frac{1}{2}\sqrt{\frac{3}{4\pi}}
\frac{\mpl}{m}~.
\end{equation}
Since the values of $m$ used in this paper are well below the
Planck scale, the stochastic effects are expected to be
negligible even for field values of a few Planck
scale~\cite{Martin:2005ir}.



\subsubsection{Exploration of the space of initial conditions}
Let us now study the space of initial values [i.e. the $(\phi_{\rr
i},\psi_{\rr i})$ plane] of the fields that lead to successful
inflation. For simplicity, we have assumed initial velocities to
be vanishing $\dot{\phi}_\mathrm{i}=\dot{\psi}_\mathrm{i}=0$ as
their effect can always be mimicked by starting in a different
point with vanishing velocities. Then for each initial conditions,
we have integrated the equations of motion and computed the field
values and the number of e-folds as a function of time. Choosing
to end simulations when inflation is violated would have not
allowed us to study trajectories where inflation is transiently
interrupted as it may happen (see Sec~\ref{sec:1fielddynamics}).
Therefore, we chose to end the numerical integration when the
trajectory is sure to be trapped by one of the two global minima,
because at that point, no more e-folds will be produced. This is
realized when the sum of the kinetic and potential energy of the
fields is equal to the height of the potential barrier between the
vacua, i.e. when
\begin{equation}
\lambda M^4=\frac{1}{2}\left(\dot\phi^2+\dot\psi^2\right)
+V(\phi,\psi).
\end{equation}

We have defined ``successful inflation'' as a period that lasts at
least for 60 e-folds\footnote{Note that the number of e-folds
required to solve the horizon problem actually depends on the
energy at which inflation is realized or the reheating
temperature~\cite{Lyth:1998xn,Liddle:2003as, Ringeval:2007am}
$$N_{\rm horizon}=62-\ln (10^{16}\mathrm{GeV}/V_{\rm end}^{1/4})-\frac{1}{3}
\ln (V_{\rm end}^{1/4}/\rho_{\rm reh}^{1/4})~.$$ Here we will
assume that inflation takes place at high energy, close to the GUT
scale.}.

Let us mention that our aim here is not to provide the best fit to
the cosmological data but to explore the space of initial
conditions that lead to sufficient inflation within the hybrid
class of models. However, notice that the COBE normalisation can
always be achieved by a re-scaling of the potential without
affecting the inflaton dynamics.

\begin{figure}
\scalebox{.22}{\includegraphics{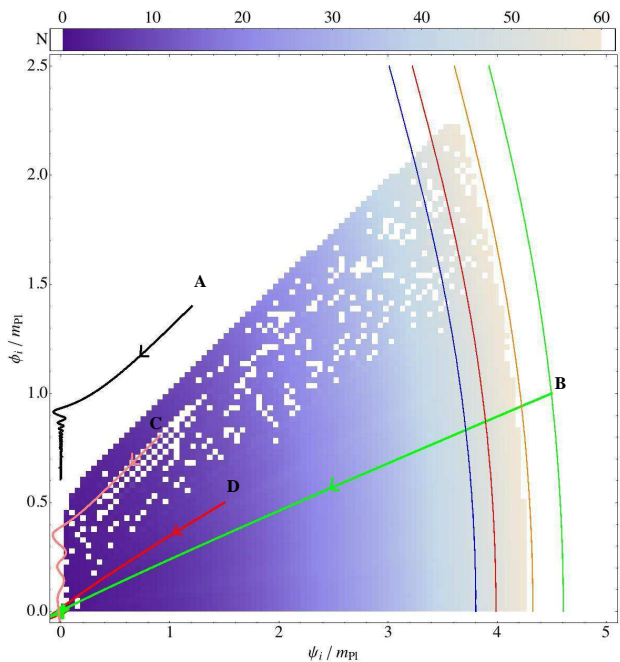}}
\caption{Grid of initial conditions leading to successful (white
regions) and unsuccessful inflation (colored region), for the
original hybrid inflation with $\lambda=\lambda'=1$, $m=10^{-6}
\mpl$ and $M=0.03 \mpl$.  The color code denotes the number of
e-folds realized. Three typical successful trajectories [in the
valley (A), radial (B), and from an isolated point (C)] are added
as well as an unsuccessful trajectory (point D).
Also plotted are the
iso-curves of $\epsone$, in the slow-roll approximation, for
$\epsone = 0.022$, $0.02$, $0.0167$ and $0.015$ (from left to
right).} \label{fig:completegrid-hybrid}
\end{figure}

\begin{center}
\begin{figure}[ht]
\scalebox{.18}{\includegraphics{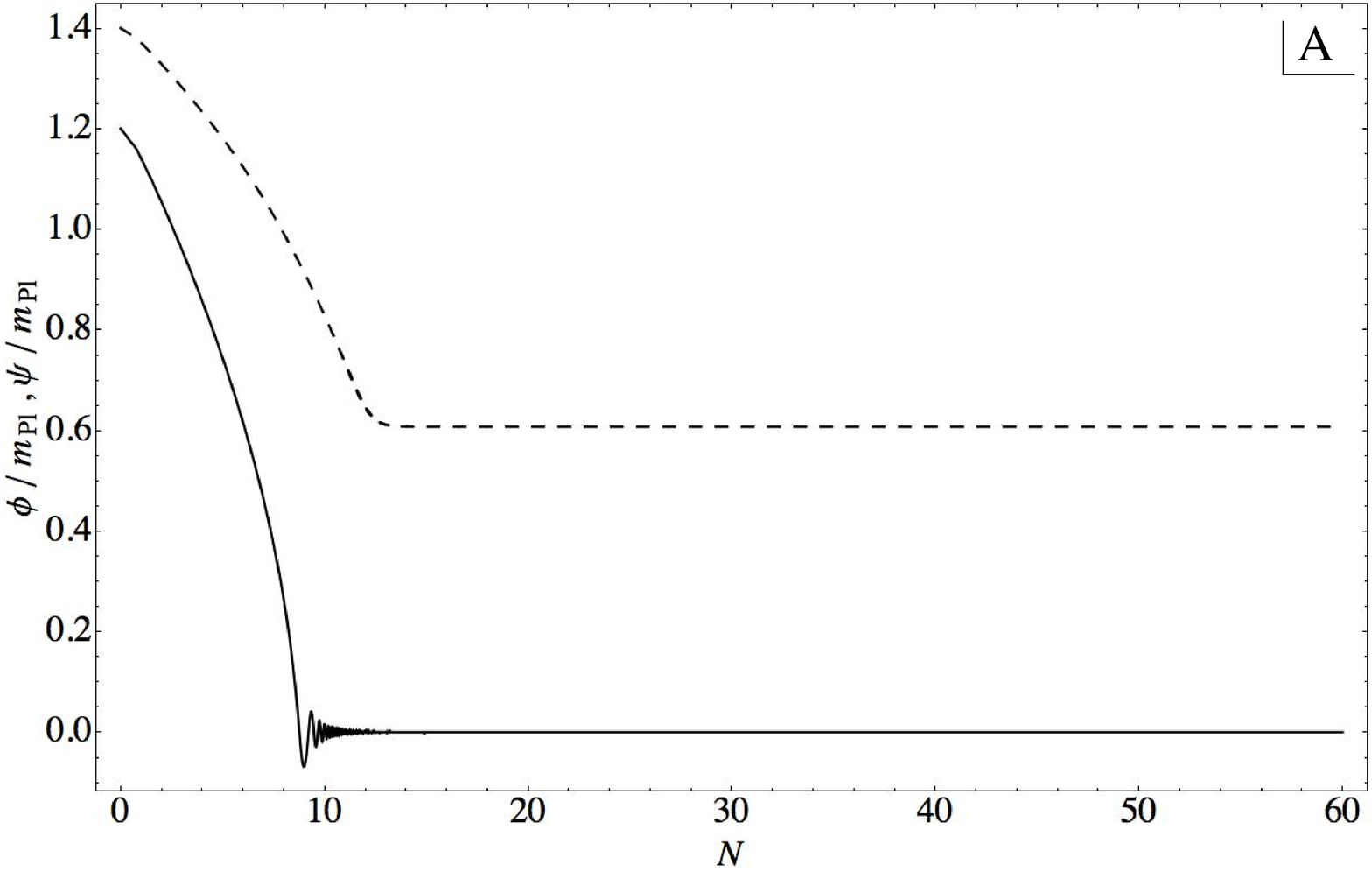}}
\scalebox{.18}{\includegraphics{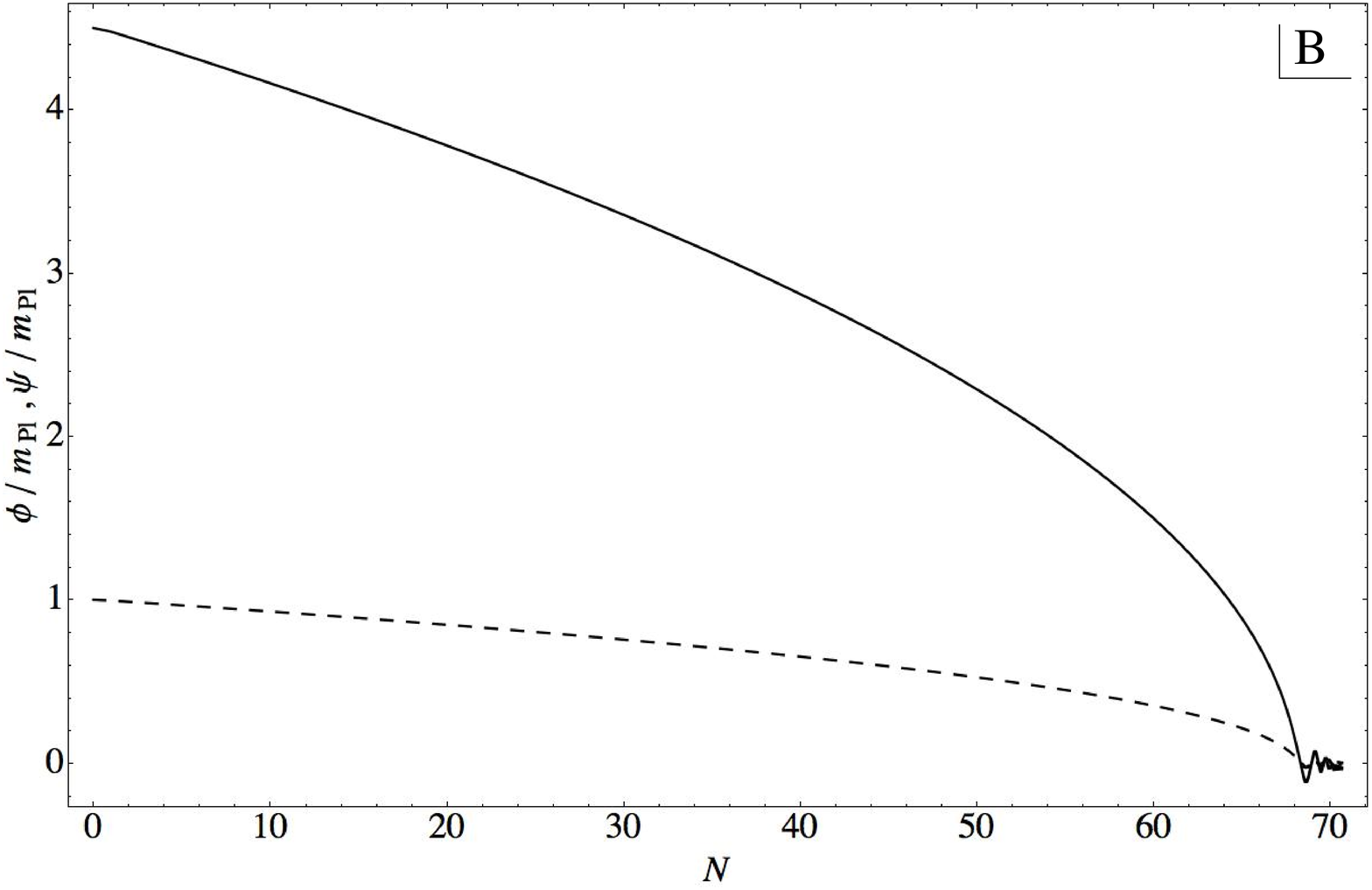}}
\scalebox{.18}{\includegraphics{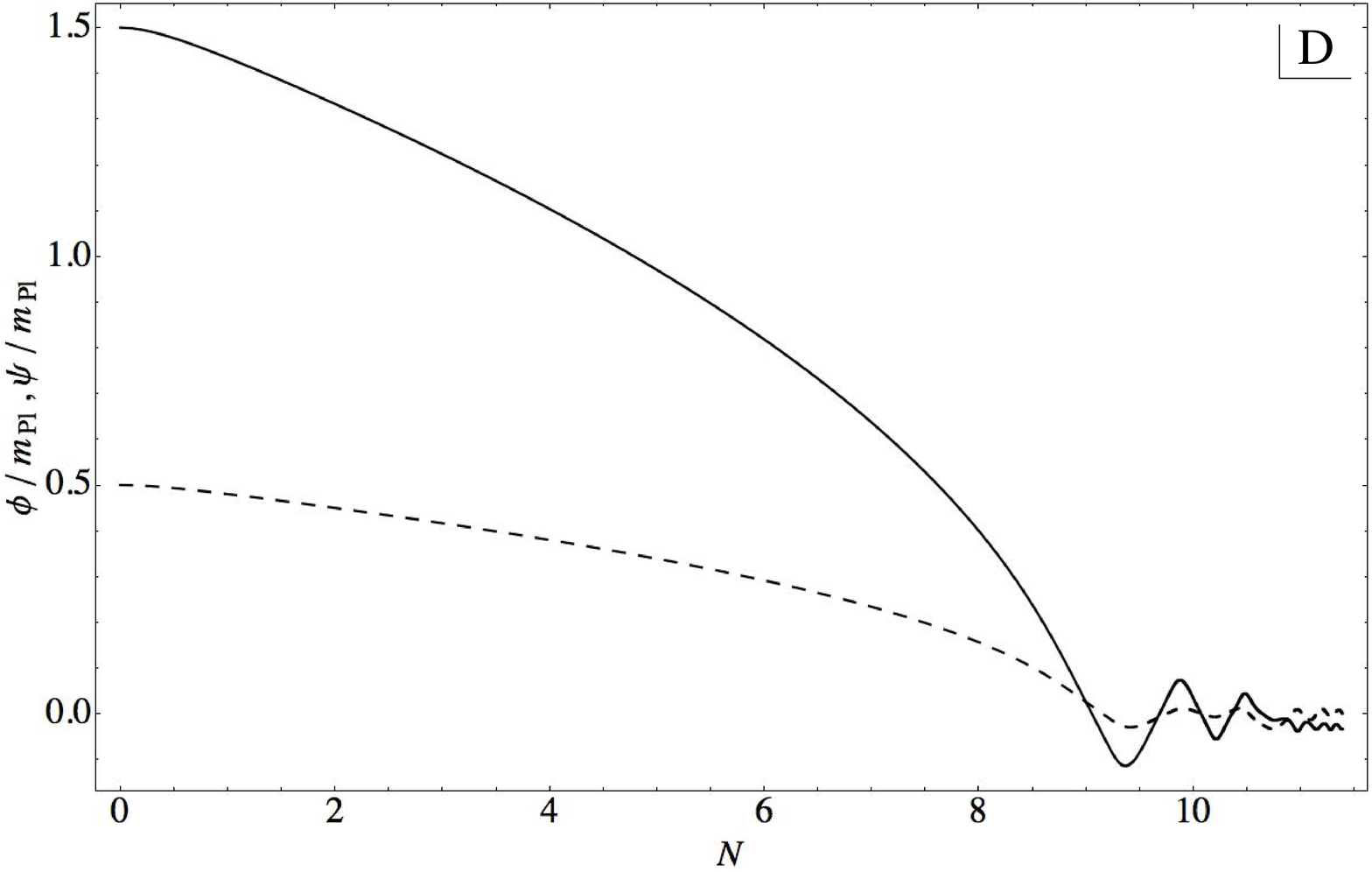}}
\caption{Evolution of the fields $\phi$ (dashed lines) and $\psi$
(plain lines) with the number of e-folds realized, for the
trajectories A, B, and D (from top to bottom) as represented in
Fig.~\ref{fig:completegrid-hybrid}. The more interesting type-C
trajectory is represented in Fig.~\ref{fig:TypeCtrajectory}
below.} \label{fig:efolds-hybrid}
\end{figure}
\end{center}

\begin{figure}
\scalebox{.2}{\includegraphics{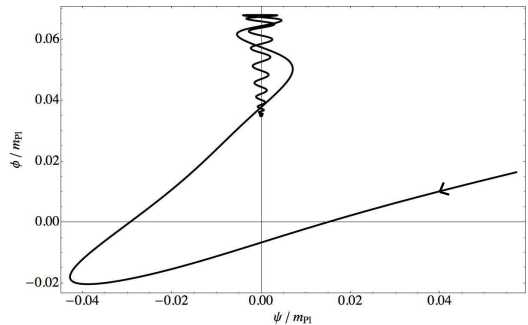}}
\caption{More detailed description of the field values during a
type-C trajectory as defined in
Fig.~\ref{fig:completegrid-hybrid}. This is a zoom of the
trajectory close to the bottom of the potential. One can notice
that the system quickly rolls down while few e-folds are produced
before ``accidently'' climbing up the valley. Then it starts a
second efficient phase of inflation like a type-A trajectory.}
\label{fig:TypeCtrajectory}
\end{figure}

In Fig.~\ref{fig:completegrid-hybrid} the grid of initial values
is presented for the original hybrid inflation model of
Eq.~(\ref{eq:potenhyb2d}). For values of parameters comparable to
those used in~\cite{Tetradis:1997kp} and \cite{Mendes:2000sq}, we
have put in evidence three types of trajectories in the fields
space to obtain successful inflation. An example of each has been
represented in Fig.~\ref{fig:completegrid-hybrid} and identified
by a letter A, B, or C whereas an example of a failed trajectory
is identified by a D.
The details of these trajectories are represented in
Fig.~\ref{fig:efolds-hybrid} where the values of the fields for
three trajectories are plotted as a function of the number of
e-folds.
A more detailed description of the more interesting type-C
trajectory is represented separately in
Fig.~\ref{fig:TypeCtrajectory}. Each trajectory is described and
explained below.\\

\paragraph{Trajectory A: along the valley}
This region of successful inflation corresponds to a narrow band
along the $\psi = 0$ line and is the standard evolution.
Trajectories are characterized first by damped oscillations around
the inflationary valley which does not produce a significant
number of e-folds. However once the oscillations are damped, the
evolution is identical to the one for the effective one-field
potential and inflation becomes extremely efficient in terms of
e-folds created. This explains the abrupt transition between the
unsuccessful and the successful type-A regions observed in
Fig.~\ref{fig:completegrid-hybrid}. Indeed, unsuccessful points
with around 10 e-folds created can be found right next to the
white successful region where $N \gg 60$. The difference between
two close points in each region is that for the successful one,
the system just has the right amount of time for the oscillations
to become damped before entering the global minimum where
inflation ends.

For larger initial values of the $\phi$ field (around and above
the Planck mass), the narrow band of successful inflation opens up
and inflation is always successful (in agreement
with~\cite{Tetradis:1997kp,Lazarides:1997vv}. In this region (at
the top of Fig.~\ref{fig:completegrid-hybrid}), it is always
possible for the oscillations to become damped and for the
efficient regime of inflation to start before the end of
inflation: \emph{the fine-tuning on the initial conditions
disappears at large values of $\phi$ for any values of $\psi$}.
This behavior is similar to the chaotic inflation model, where
\cite{Linde:1983gd} super-planckian values are necessary to have a
long enough inflationary phase.

By comparing the time necessary for the expansion to damp the
oscillations and the time taken by the inflaton to reach the
critical point of instability, an analytical approximation of the
width $\psi_{\rr w}$ of the narrow successful band has been
proposed in~\cite{Tetradis:1997kp},
\begin{equation}\label{eq:largeurvallee}
\psi_{\rr w} \simeq \sqrt{\frac{3\lambda\pi^3}{4\lambda'}} \frac{M^2}{\mpl}~.
\end{equation}
For the parameter values of the
Fig.~\ref{fig:completegrid-hybrid}, $\psi_{\rr w} \sim 4\times 10^{-3} \mpl$.
This provides a good fit of the width of the inflationary valley
at small $\phi \ll \mpl$. This successful band is so thin that
quantum fluctuations would have an amplitude large enough to shift
the field $\psi$ outside the successful
band~\cite{Tetradis:1997kp}.  For larger initial values of $\phi$,
it is also possible to provide an analytical fit of the limit
successful/unusuccessful. Fig.~\ref{fig:completegrid-hybrid}
suggests that the limit $\phi_{\rm lim}(\psi)$ is a linear
function.  From a given set of initial conditions $(\phi_\ui,
\psi_\ui)$, the total number of e-folds generated depends almost
only on the value $\phi=\phi_\uhit$ at which the oscillations in
$\psi$ become damped and the slow-roll starts in the valley. The
reason is that a type-A trajectory rolls faster before $\phi_\uhit$
and thus doesn't generate many e-folds before the valley. As a
consequence, the limit between successful and unsuccessful regions
necessarily follows the unique trajectory for which $\phi_\uhit$
becomes large enough to generate exactly $60$ e-folds by slow-roll
in the valley. As a result, using the slow-roll approximation, the
slope of the limit is simply given by the gradient of the
potential
\begin{equation}\label{eq:pentelimit}
\alpha = \frac{\partial V(\phi,\psi)/\partial \phi} {\partial
V(\phi,\psi)/\partial \psi}\simeq \frac{\lambda'\phi\psi}
{\lambda\psi^2+\lambda'\phi^2}~,
\end{equation}
where the approximated expression is valid when mass parameters
are small  $\psi \gg \max(m,m/\lambda')$. Given one point of the
transition line, for example $(1,1)$, we can check that the slope
of the limit is $\alpha \simeq 0.5$ for the parameters of
Fig.~\ref{fig:completegrid-hybrid}.\\

\paragraph{Trajectory B: radial}
Enlarging the space of initial conditions to super-planckian values
shows another region where successful inflation is automatic. It is
observed for super-planckian initial values of the auxiliary field
$\psi$ beyond a few Planck mass,  in a way reminiscent to the chaotic
scenario. In this case, the trajectory is called radial and the $60$
e-folds are realized mostly before reaching the valley or the global
minima.

From the $\phi$-axis to larger values of $\psi_i$, the number of
e-folds realized increases slowly (see
Fig.~\ref{fig:completegrid-hybrid}). Therefore, this limit between
the two regions is smooth unlike the limit with A-type
trajectories described at the previous paragraph. Increasing
$\phi_i$, the critical value of $\psi_i$ leading to enough
inflation decreases slowly, because inflation is radial and the
trajectory longer. To describe this limit more precisely, we have
plotted the iso-curves of $\epsone$ in
Fig.~\ref{fig:completegrid-hybrid}) in the two-fields slow-roll
approximation. We can see that this limit follows one of these
iso-curves, namely $\epsone\simeq 0.0167$. This observation can be
understood using a kinematic analogy~\cite{Ringeval:2007am} as
long as $\epstwo$ is negligible. This critical value of $\epsone$
can be computed analytically, by studying the easiest trajectory
of this kind at $\phi_i = 0 $. In this case, the effective
potential is dominated by $\lambda \psi^4 $, and the critical
$\psi_i$ is obtained by requiring a phase of inflation of exactly
$N_{\rr{suc}}=60$ e-folds. We find
\begin{equation}
\psi_{\rr {ic}} = \sqrt{\frac {\mpl ^2}{\pi} \,N_{\rr{suc}} }
\approx 4.37 \mpl.
\end{equation}
At this value, the corresponding first Hubble-flow parameter
$\epsilon_{\rr {1c}}$ reads
\begin{equation}
{\epsone}_{\rr c} \simeq \frac {1}{N_{\rr{suc}}} \approx 0.0167.
\end{equation}

\paragraph{Trajectory C and D: isolated successful points and
unsuccessful points.}
Previous works~\cite{Tetradis:1997kp,Mendes:2000sq} pointed out
the presence of unexplained successful isolated points in the
central unsuccessful region. In this paragraph, we justify their
existence, study their properties and quantify the area they
occupy.

Let us first describe the D-type trajectories that are
unsuccessful. As shown in Fig.~\ref{fig:efolds-hybrid}, in these
cases, the system quickly rolls down the potential to one of the
global minima of the potential during which only a few e-folds are
created. What is then the difference between the D-type and the
C-type trajectories  plotted in Fig.~\ref{fig:TypeCtrajectory} ?
The fields roll towards the bottom of the potential with
sufficient kinetic energy and, after some oscillations close to
the bottom of the potential, the momentum is ``by chance''
oriented toward the inflationary valley. Thus the system goes up
the valley until it looses its kinetic energy and then starts
slow-rolling back down the same valley producing inflation with a
large number of e-folds.
Note that there are more of these points in a band under the
limit of type-A trajectories. This is because, at higher
$\phi_\ui$, there are more chances to find a trajectory where the
momentum at the bottom of the potential is oriented toward the
inflationary valley.

\begin{figure}
\scalebox{.19}{\includegraphics{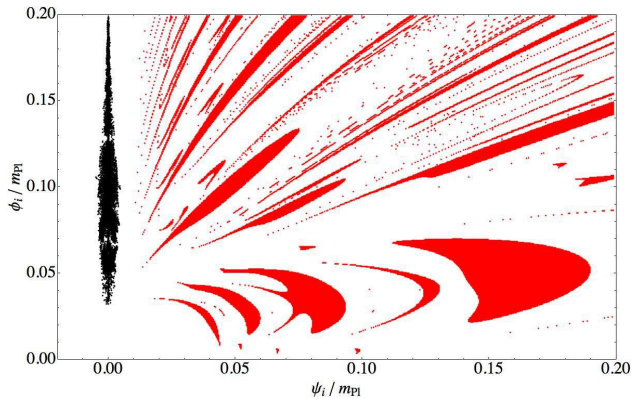}}
\caption{Structure of the successful ``anamorphosis points'' (in grey)
together with their images (in black) defined by the point
of the trajectory at which the velocities of the fields vanish.
The structure in grey can be seen as the ``anamorphosis'' of
the patterns of successful inflation in black. In this analogy the
trajectories of the light on the optic device in order to create a
meaningful image are replaced by the trajectories of the system in
field space to create a meaningful image (in the valley) from the
apparently senseless grey patterns. This is obtained for
$M=0.03 \mpl$, $m=10^{-6}\mpl$,
$\lambda=\lambda'=1$.}\label{fig:anamorphosis}
\end{figure}

High resolution grids and zooms on peculiar regions of
Fig.~\ref{fig:completegrid-hybrid} show that these apparently
random isolated points form actually a complex structure. Some of
it, for small initial conditions, is visible in
Fig.~\ref{fig:anamorphosis}.
The points are organized in long thin lines, or croissants. The
points that seem isolated actually belong to structures that a
better resolution would show continuous. Some of our biggest
structures can be identified also in~\cite{Tetradis:1997kp} but
are not recovered in~\cite{Mendes:2000sq} where only isolated
points were found. This may be explained by the need of a higher
resolution to resolve the structures. A detailed analysis of
trajectories shows that for each continuous successful region
corresponds a unique number of crossings the $\phi = 0 $ axis by
the trajectory before climbing up and going back down the
inflationary valley along the $\psi=0$ direction.

For each of these type-C trajectories, we can identify the point
(that we will call the ``image'') on the inflationary valley at
which the velocities of the fields become (quasi)null. We show the
robustness of the previous description of the type C trajectories,
by observing that all these images are in the successful narrow
band responsible for the type-A trajectories. More precisely, the
images obtained populate \emph{exactly} the narrow band of width
$\psi_w$ as described in Eq.~(\ref{eq:largeurvallee}). As a
conclusion each successful point in the unsuccessful region
corresponds to a point in the narrow successful band. The
identification between the isolated points and their images in the
inflationary valley is represented in Fig.~\ref{fig:anamorphosis},
when restricting ourselves positive initial field values. Using
the analogy with optical \emph{anamorphosis}, we can say that the
observed structures of type-C initial conditions generates by
anamorphosis the successful narrow band around the inflationary
valley. In this analogy, the potential plays the role of the
optical instrument used to create the meaningful image. The
trajectories of the light rays on the optic device are then
replaced by the field trajectories to create a meaningful image
(in the valley) from the apparently senseless patterns of
successful initial conditions.

Let us elaborate a little more on the properties of the images in
Fig.~\ref{fig:anamorphosis}. Since the potential is invariant
under $\phi\rightarrow -\phi$, there exist two different
inflationary valleys, one going toward $\phi>0$ and one going
$\phi<0$. Some of these type-C initial conditions give rise to
inflation thanks to the first valley when the others will realize
inflation in the second. Obviously the two situations are
equivalent and symmetric, just like the value of the initial
conditions that could be taken in the negative planes (with
$\phi_i<0$ or $\psi_i<0$ or both). It is clear that these
additional planes contain identical patterns and some of these
initial conditions would populate as well the inflationary valley
represented above. Moreover, the set of images in the valley
represented in Fig.~\ref{fig:anamorphosis} is not of constant
width as it should be for another reason. The initial conditions
represented are restrained below $\Mpl$ but the structures
observed continue at larger values. The trajectories starting with
these larger values of $\phi_i$ would have the momentum to climb
up the valley more and populate the higher part of the set of
images.

\subsubsection{Dependencies on the parameters}
The grid of initial conditions, and therefore the proportion of
successful points in a given range of initial values naturally
depend on the values of the parameters of the potential. Three
physical quantities are of interest to study these evolutions: the
width of the inflationary valley, proportional to
$M\sqrt{\lambda/\lambda'}$, its length
controlled by the critical value $\phi_\mathrm{c}=M\sqrt{\lambda/
\lambda'}$, the depth of the global minima of the potential given
by $V_0\propto \lambda M^4$, and the gradient of the
potential $\alpha \simeq \lambda'/(\lambda+\lambda')$.\\

\begin{figure}[h!]
\scalebox{.45}{\includegraphics{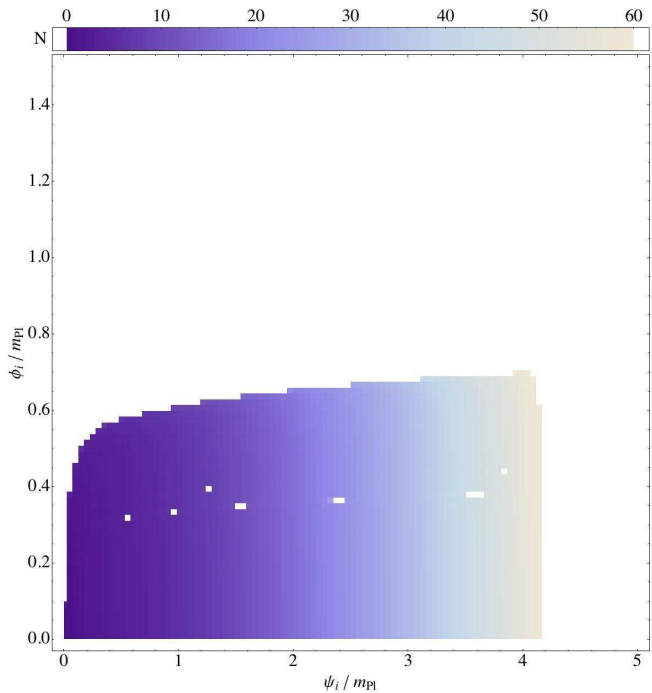}}
\caption{Grid of initial conditions, for hybrid potential with
$M=0.03 \mpl$, $m=10^{-6} \mpl$, $\lambda=1$, $\lambda ' = 0.1 $.}
\label{fig:hybridVaryLambda'}
\end{figure}

\paragraph{Evolution of the limit of A-type trajectories}
At small $\phi$, a smaller $\lambda$ induces a narrower
inflationary valley, and therefore fewer successful initial
conditions. At large $\phi$, the slope $\alpha$ in
Eq.~(\ref{eq:pentelimit}), is mostly a function of the coupling
constants. For a smaller value of $\lambda$, the slope of the
limit increases.
For smaller value of $\lambda'$, the slope of the limit is reduced
as represented on Fig.~\ref{fig:hybridVaryLambda'}. This effect is
due to the potential now dominated by the $\lambda\psi^4$ term,
depending less on $\phi$. Thus the velocity in the $\psi$
direction is enhanced compared to the $\phi$ one.\\

\begin{figure}
\scalebox{.4}{\includegraphics{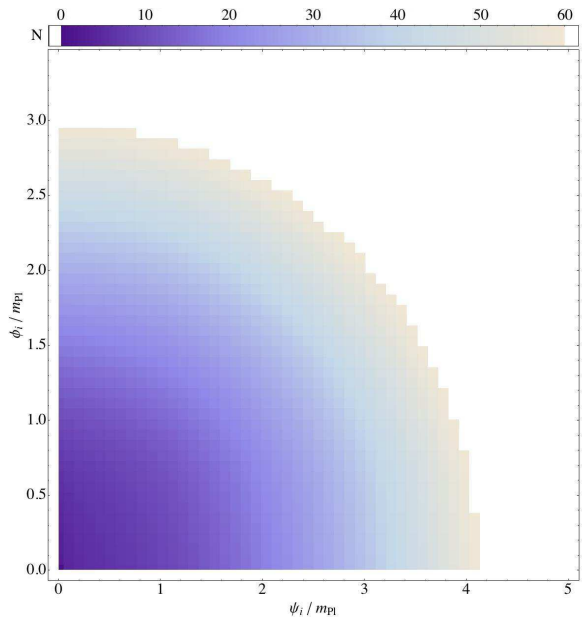}}
\caption{Grid of initial conditions and example trajectories for
the hybrid model, with $m=M=10^{-3} \mpl$, $\lambda=1$, $\lambda'=
10^{-2}$.}\label{fig:hybrid_redspectrum}
\end{figure}

As long as the mass squared $m^2$ is subdominant compared to
$\lambda'\psi^2$, its variation don't affect the properties of the
initial condition plane. Increasing the mass $m$ above this limit
increases the velocity in the $\phi$ direction and tends to spoil
the slow-roll evolution in the inflationary valley. As already
described in the section~\ref{sec:1fielddynamics}, this violation
of the slow-roll conditions in the valley imposes for inflation to
occur in the large field phase. In the space of initial
conditions, the narrow successful band then disappear together
with the type-C trajectories. Finally the unsuccessful region
takes an elliptic form as represented in
Fig.~\ref{fig:hybrid_redspectrum}, with a smooth transition
between successful and unsuccessful regions. The model becomes
comparable to the sum of two chaotic inflation models and we
recover the feature of these models: it is almost unavoidable to
have super-planckian initial values of the fields to realize a
sufficiently long inflation.

\paragraph{Evolution of the amount of C-type trajectories}
A similar explanation can be given to justify the absence of
isolated points for small values of $\lambda'$ (see
Fig.~\ref{fig:hybridVaryLambda'}). Even though the width of the
valley is larger, the potential is then dominated by the $\lambda
\psi^4$ term and the $\phi$-component of the velocity becomes
small. Thus the chances for the system to climb up the valley are
suppressed. For larger values of the parameters $M$ and $\lambda$,
these isolated points also disappear, because the two minima of
the potential are deeper, and there is a larger chance for the
system to get trapped in them without climbing up the inflationary
valley.
These results are summarized in Tab.~\ref{tab:anamorphhyborigin}
below.\\

\paragraph{Quantification of successful initial conditions}
We end this section by quantifying what proportion of the initial
condition space give rise to inflation for hybrid inflation, for
various values of the parameters, including the proportion of
points in the anamorphosis. Our results are represented in
Tab.~\ref{tab:anamorphhyborigin}, where the quantification is
first made restricting the amplitude of the fields below the
reduced Planck mass.
\begin{table*}
\begin{center}
\begin{tabular}{|c|c|c|c|c|}
\hline Model & Values of parameters & Successful points (\%) &
Anamorphosis points (\%) & Figure \\\hline\hline
Hybrid & $M=0.03 \mpl$, $m=10^{-6} \mpl$, $\lambda=\lambda'=1$ & 17.4  & 14.8 & \ref{fig:completegrid-hybrid} \\
Hybrid & $M = 0.06 \mpl$, $m=10^{-6}$, $\lambda=1$, $\lambda'=1$ & 11.3 & 5.5 &\\
Hybrid & $M=0.03 \mpl$, $m=10^{-5} \mpl$, $\lambda=\lambda'=1$ & 17.4 & 14.8 &\\
Hybrid & $M=0.03 \mpl$, $m=10^{-6} \mpl$, $\lambda=0.1$, $\lambda'=1$ & 15.5 & 14.1 & \\
Hybrid & $M=0.03 \mpl$, $m=10^{-6} \mpl$, $\lambda=1$, $\lambda'=0.1 $ & 2.8 & $<0.1$ &\ref{fig:hybridVaryLambda'}\\
Hybrid & $M = m =10^{-3} \mpl$, $\lambda=1$, $\lambda'=10^{-2}$ &
0 & 0 & \ref{fig:hybrid_redspectrum}\\\hline
\end{tabular}
\caption{Percentage of successful points in grids of initial
conditions, for different values of parameters, when restricting to
$\phi_\ui,\psi_\ui\leq \Mpl$. The third column represents the area
of the whole successful initial condition parameter space over the
total surface. The fourth column represents the surface of the
successful space only located in isolated points, over the total
surface. This allows to visualize the importance of these isolated
points. For several of these sets of values for the potential
parameters, the grid of initial conditions is represented in the
body of the paper. When it is the case, the number of the figure
is given in column 5.} \label{tab:anamorphhyborigin}
\end{center}
\end{table*}
From this table, we can see that unless $\lambda'$ is very small,
or $M$ is close to planckian values, the hybrid model possesses
about $15\%$ of initial conditions that leads to successful
inflation. For this percentage to be translated into a probability
of realizing inflation, one would need a measure in the
probability space. If this measure was to be flat, \emph{the
successful initial conditions should not be considered as
fine-tuned but simply sub-dominant when fields are restricted to
sub-planckian values}.

From Fig.~\ref{fig:completegrid-hybrid}, it is obvious that if we
don't require that the fields are smaller than the reduced Planck
mass, the proportion of successful initial conditions will tend
toward $100\%$. Therefore, we have also realized the same
quantification with the requirement $\phi_i, \psi_i \leq 5 \mpl$
and found that the percentage of successful initial conditions
raise to $72\%$ for the parameter values of
Fig.~\ref{fig:completegrid-hybrid}.

\section{Initial conditions for extended models of hybrid inflation}\label{sec:othermodels}
In this section, we will study the properties of initial
conditions leading to successful inflation for three hybrid-type
models of inflation and study how generic the properties observed
for the original model are. The models are the ``smooth'', and
``shifted'' hybrid inflation both in global SUSY and SUGRA, and
radion inflation.

\subsection{Motivations for smooth and shifted hybrid inflation}

Following the original inflation model, a supersymmetric
formulation, the F-term hybrid inflation, has been proposed
by~\cite{Dvali:1994ms}. In this case, the inflaton field $\phi$ is replaced by
a superfield $S$, and the Higgs field $\psi$ is replaced by a pair
of Higgs superfields $\bar{\Phi},\Phi$ non-trivially charged under
a symmetry group\footnote{They are assumed to belong to two
complex conjugate representations.} $G$ whereas $S$ is assumed to be a
gauge singlet of $G$. The only superpotential, invariant under
$G$ and under an R-symmetry\footnote{This R-symmetry is a $U(1)$
symmetry under which $\Phi$ and $\bar{\Phi}$ have opposite charges
and $S$ and $W$ have identical charges.} and containing only
renormalizable terms reads~\cite{Dvali:1994ms}
\begin{equation}\label{eq:ftermsuperpot}
W^{\mathrm{F}} = \kappa S (\Phi_+ \Phi_-  - M^2)~.
\end{equation}
It gives rise to a scalar potential similar to
Eq.~(\ref{eq:potenhyb2d}), the coupling constants $\lambda$ and
$\lambda'$ being replaced by $\kappa$. This potential possesses
the same features with the inflationary valley along
$\bar{\Phi},\Phi=0$, this valley being destabilized when one of
the superfields $\bar{\Phi},\Phi$ becomes tachyonic. The field
develops a non-vanishing vev which leads to the breaking of
$G$. Topological defects can be produced during this breaking,
depending on $G$. They can be cosmic strings~\cite{Jeannerot:2003qv}
which would be in agreement with the most recent CMB
data~\cite{Rocher:2004et,Jeannerot:2005mc,Fraisse:2006xc},
provided that their effect on the CMB is subdominant~\cite{Bevis:2007gh}.
But they could also be monopoles or domain walls and then be in
contradiction with observations~\cite{Vilenkin:1994}.

To be able to implement hybrid inflation at any symmetry breaking,
it has been proposed two extensions of the F-term
model: the smooth~\cite{Lazarides:1995vr} and the
shifted~\cite{Jeannerot:2000sv} hybrid inflation. They are both
based on the idea of shifting the inflationary valley away from
$\psi=0$. As a consequence the symmetry $G$ is broken \emph{during
or before} inflation, and thus any topological defect formed
during this breaking are diluted away by inflation. This is
achieved by introducing non-renormalizable terms in the
potential~\cite{Lazarides:1995vr,Jeannerot:2000sv} and imposing an
additional discrete symmetry for the
superpotential~\cite{Lazarides:1995vr}.

As detailed in the introduction, if these models are considered
realistic, that is if the scalar potential is assumed to be
originated from SUSY or SUGRA, it is not safe to consider
super-planckian fields. It can also be safe to study these models
beyond super-planckian fields if they originate from other
frameworks where non-renormalizable corrections are controlled or
prevented.

\subsection{Smooth Inflation}
\subsubsection{The potential in SUSY}
Smooth inflation has been introduced by Lazarides and
Panagiotakopoulos~\cite{Lazarides:1995vr}.  It assumes that the
superpotential is invariant under a $Z_2$ symmetry under which
$\Phi\bar{\Phi}\rightarrow -\Phi\bar{\Phi}$. This forbids the
first term in the F-term superpotential of Eq.~(\ref{eq:ftermsuperpot})
but allows for one non-renormalizable term\footnote{Note that our
choice of setting the renormalization scale to the reduced Planck
mass is arbitrary. In general, we can write $W^{\mathrm{sm}}=\kappa S
\left[-M^2+(\bar{\Phi}\Phi)^2/\Lambda^2\right]$, $\Lambda$
corresponding to the scale of new physics.}~\cite{Lazarides:1995vr}
\begin{equation}
W^{\mathrm{sm}}=\kappa S \left[-M^2+\frac{(\bar{\Phi}
\Phi)^2}{\Mpl^2}\right]~.
\end{equation}
In the context of global supersymmetry, the scalar potential
reads~\cite{Lazarides:1995vr}
\begin{equation}
\begin{aligned}
V^{\rr{sm}}(S,\Phi,\bar{\Phi}) & = \kappa^2 \left| -M^2 + \frac{(\bar
\Phi \Phi)^2 }{\Mpl^2 } \right|^2\\
& + 4\kappa^2 | S| ^2 \frac {|\Phi|^2 |\bar \Phi|^2  }{\Mpl^4}
\left( |\Phi|^2 + |\bar \Phi|^2  \right),
\end{aligned}
\end{equation}
where we denote by the same letter the superfields and their
scalar components.
Two real scalar fields $\phi$ and $\psi$ can be defined as the
relevant components of the $S$, $\Phi$, $\bar \Phi$ fields such
that the fields are canonically normalized
\begin{equation}\label{eq:defininflatwaterfall}
\phi\equiv \sqrt{2} \mathrm{Re}(S)~,\quad
\psi\equiv 2\mathrm{Re}(\Phi)=2\mathrm{Re}(\bar{\Phi})~,
\end{equation}
and the potential becomes~\cite{Lazarides:1995vr}
\begin{equation}
V^{\rr{sm}}(\phi,\psi)=\kappa^2\left(M^2-\frac{\psi^4}{16\Mpl^2}
\right)^2  + \kappa^2 \phi^2 \frac{\psi^6}{16\Mpl^4}\,.
\end{equation}
This potential contains a flat direction $\psi=0$, but it is a
local maximum. The global minima are obtained for non-vanishing
values of $\psi$: they define two distinct inflationary valleys,
along
\begin{equation}
\psi=\pm\sqrt{-6\phi^2+6\sqrt{\phi^4+\frac{4}{9}M^2 \Mpl^2}}~.
\end{equation}
Note that these inflationary valleys progressively shift away from
$\psi=0$ as $\phi$ evolves towards $0$.

\subsubsection{Space of initial conditions}

In a previous study by Lazarides et al.~\cite{Lazarides:1996rk},
an exploration of the space of initial conditions leading to
sufficient inflation was performed, with a low resolution.  This
exploration led to a conclusion opposite to the one found for the
non-supersymmetric hybrid inflation model: most of the space
was found to be successful. Therefore, smooth hybrid inflation seems
a good laboratory to test the validity of the results we found at
the previous section. We performed the exploration of the space of
initial conditions, for a higher resolution, and for a larger
range of initial field values and parameter values. Imposing
$\phi_i,\psi_i \leq \Mpl$, we computed the proportion of successful
initial conditions and the proportion of isolated successful points
away from the inflationary valleys.

\begin{center}
\begin{figure}[h!]
\scalebox{.45}{\includegraphics{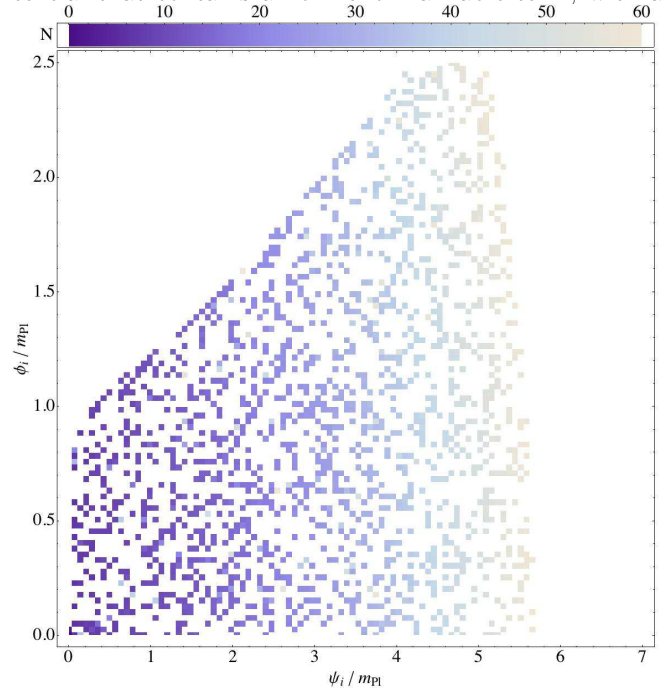}}
\caption{Grid of initial conditions for smooth inflation, using
the values of the parameters of \cite{Lazarides:1996rk}: $\kappa
\simeq 10$, $M\simeq 2.3\times 10^{-5} \mpl$.}
\label{fig:smoothVaryAllLazarides}
\end{figure}
\end{center}

Our study, also extended to super-planckian values of the fields,
always reveals a structure similar to that of the original model.
We observe (see for e.g. Fig.~\ref{fig:smoothVaryAllLazarides})
a narrow band of
fine-tuned successful initial conditions along $\psi=0$, a
triangular unsuccessful region, and successful areas for large
initial values of one or both of the fields. Anamorphosis is also
present, leading to isolated successful patterns in the
unsuccessful region. For the values of the parameters quoted
in Ref.~\cite{Lazarides:1996rk}, that is with a mass scale of
order $10^{-5} \mpl$, they occupy most of the space of
initial condition as shown on Fig.~\ref{fig:smoothVaryAllLazarides}.
We find almost $80\%$ of initial conditions below the reduced
Planck mass to be successful.

We have also studied how this grid evolves with the parameters of
the potential. We first observe that the amount of successful
initial conditions is independent of the coupling constant
$\kappa$ (it only scales the potential or the CMB spectrum), but
only depends on the mass scale $M$. This analysis shows a strong
dependency with the value of $M$, the amount of successful initial
conditions ranging from $15\%$ to almost $80\%$ when $M$ ranges
from $10^{-2}$ and $10^{-5}$. For $M$ below the GUT scale,
$10^{16}$ GeV, the quantification of successful initial conditions
is larger than $50\%$, providing a good mechanism to produce
inflation without fine-tuning of initial conditions. As a
conclusion, we confirm the qualitative results
of~\cite{Lazarides:1996rk}, and we note that they depend on the
values of potential parameters. We note also that most of the
successful initial conditions are isolated, that is located
outside of the inflationary valleys: they form an anamorphosis
like in the hybrid inflation model. These results are summarized
in the Tab.~\ref{tab:anamorph} at the end of this section.

\subsubsection{Supergravity corrections}
The smooth hybrid inflation is based on a superpotential that
contains a non-renormalizable term, with a cutoff scale chosen at
the reduced Planck mass. In addition, in our study we consider
field values that are non-negligible compared to $\Mpl$, sometimes
above. Therefore, to extend the domain of validity of the model,
supergravity corrections (introducing corrections proportional to
negative powers of $\Mpl$) should be taken into account. We remind
the reader that outside of the domain of validity of the model
(whether in SUSY or in SUGRA), the model is still studied but
considered as an effective model derived from some frameworks in
which super-planckian field values can safely be considered (see
introduction).

Assuming supergravity with a minimal K\"ahler potential,
\begin{equation}
K=K_{\rm min}=|\Phi|^2+|\bar{\Phi}|^2+|S|^2,
\end{equation}
the scalar potential reads,
\begin{widetext}
\begin{equation}
\begin{aligned}
&V_{\rm SUGRA}^{\rr{sm}}(S,\Phi,\bar{\Phi})  = \kappa^2\mathrm{Exp}
\left[\frac{K_{\rm min}}{\Mpl^2}\right]\left\{ \left| \frac{(\bar
\Phi \Phi)^2 }{\Mpl^2 } -M^2 \right|^2\left(1-\frac{|S|^2}{\Mpl^2}
+\frac{|S|^4}{\Mpl^4}\right)\right.\\
+&\left. \frac {|S|^2}{\Mpl^4}\left[\left(\left| \frac{(\bar
\Phi \Phi)^2 }{\Mpl^2 }-M^2  \right|^2 + 4 |\Phi|^2 |\bar \Phi|^2 \right)
\left( |\Phi|^2 + |\bar \Phi|^2 \right)+ 4\Phi^2
\bar{\Phi}^2\left( \frac{(\bar
\Phi^* \Phi^*)^2}{\Mpl^2 }-M^2\right)+c.c.\right]\right\}~.
\end{aligned}
\end{equation}
\end{widetext}
This potential is in agreement with~\cite{Yamaguchi:2004tn}, though
all terms have here been kept since in our study, fields are not
necessarily small compared to the Planck mass. We define again the
inflaton and waterfall fields like in
Eq.~(\ref{eq:defininflatwaterfall}), and we obtain the full potential
in SUGRA,
\begin{equation}
\begin{split}
V_{\rm SUGRA}^{\rr{sm}}&(\phi,\psi)=\kappa^2\mathrm{Exp}\left[\frac{\phi^2+\psi^2}
{2\Mpl^2}\right]\left\{\left(M^2-\frac{\psi^4}{16\Mpl^2}\right)^2\right.\\
&\times\left(1-\frac{\phi^2}{2\Mpl^2}+\frac{\phi^4}{4\Mpl^4}
+\frac{\phi^2\psi^2}{4\Mpl^4}\right)  \\
&\left.+ \frac{\phi^2\psi^6}{16\Mpl^4}-\frac{M^2\phi^2\psi^4}{4\Mpl^4}
+\frac{\phi^2\psi^8}{64\Mpl^6} \right\}~.
\end{split}
\end{equation}

SUGRA corrections induce a steeper potential in the large field regime. We
have studied for this last potential the space of initial condition
leading to enough inflation and compared the results to the SUSY case. We
observe two properties of the space of initial conditions.
First, at low initial field values, the initial condition space is mostly
unchanged because the correction are small. In particular, the patterns of
isolated successful initial conditions still exist and are as numerous.
This can be understood because even if SUGRA corrections induce higher
velocities for the fields, the anamorphosis mechanism can take place just
as easily: the fields
fast-roll down the potential, oscillate more around the bottom and sometimes
climb up one of the inflationary valleys. When the fields slow-roll back down
the valleys, enough inflation is generated. We note that the probabilities
to realize inflation this way are higher in SUGRA than in SUSY and the total
and can be as high as $70\%$ for small values of $M$.
Second, the steepness of the potential induces a violation of slow-roll for
radial trajectories leading first to a smaller probability to realize inflation
directly starting in the inflationary valleys because they become more narrow.
Second we don't observe automatic successful inflation at large
super-planckian initial field values and similar patterns of isolated points
are observed instead.

\subsection{Shifted Inflation}

\subsubsection{The potential}

\begin{center}
\begin{figure}
\begin{center}
\scalebox{.8}{\includegraphics{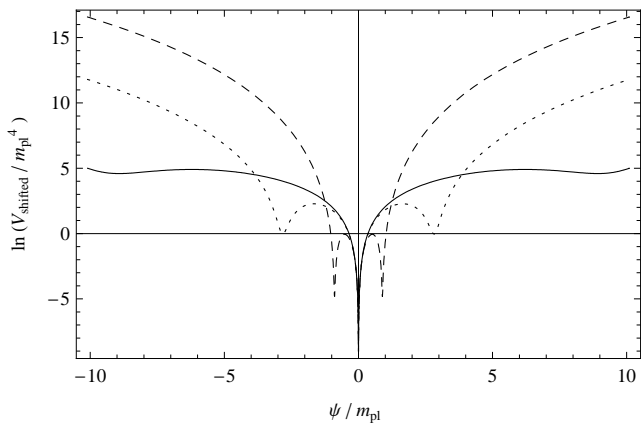}}
\caption{Cut of the logarithm of the shifted potential $V^{\rm sh}(\phi,
\psi)$, at $\phi=2 \mpl$, for $M=0.1 \mpl$, $\kappa=1$, and $\beta = 10^{-3}
\mpl^{-2} $ (plain line), $\beta = 10^{-2} \mpl^{-2}$ (dotted line),
$\beta = 10^{-1} \mpl^{-2}$ (dashed line). Notice the appearance of
multiple inflationary valleys, whose number and positions depend
on the parameter $\beta$. They also depend on the value of $\phi$.}
\label{fig:shifted_potential_cut}
\end{center}
\end{figure}
\end{center}

The shifted inflation model, proposed by Jeannerot et
al.~\cite{Jeannerot:2000sv}, is similar to the smooth inflation
model, but the additional $Z_2$ symmetry of smooth inflation is
not imposed anymore.
Thus the superpotential reads
\begin{equation}
W^{\rr{sh}} = \kappa S \left(- M ^2 + \bar{\Phi}\Phi - \beta \frac{(\bar
\Phi \Phi)^2}{\Mpl^2}\right)~.
\end{equation}
This gives rise to the following F-terms contributions to the
scalar potential, in the context of global
supersymmetry
\begin{equation}
\begin{aligned}
V^{\rr {sh}}&(S,\Phi,\bar{\Phi}) =\kappa^2\left\{ \left| -M^2 + \bar{\Phi} \Phi - \beta
\frac{(\bar{\Phi} \Phi)^2}{\Mpl^2} \right|^2 \right.\\
&\left.+ \left| S \right|^2 \left( \left|  \bar \Phi \right|^2 +
\left| \Phi \right|^2\right) \left| 1 - 2
\beta \frac{\bar{\Phi} \Phi}{\Mpl^2} \right|^2\right\},
\end{aligned}
\end{equation}
where we have used the same letter to denote the superfields and
their scalar component. We can define the relevant fields
$\bar \psi$ and $\psi$ as the components of $\bar \Phi$ and $\Phi$
that generate the breaking of the group $G$. We can define the
inflaton and waterfall fields like in
Eq.~(\ref{eq:defininflatwaterfall}) so as to vanish the D-term
contributions to the potential and to have canonical kinetic terms.
The effective scalar potential then becomes~\cite{Jeannerot:2000sv},
\begin{equation}\label{}
\begin{aligned}
V^{\rm sh}(\phi,\psi) = \kappa^2 & \left( \frac{\psi^2}{4}-M^2-
\beta\frac{\psi^4}{16\Mpl^2} \right)^2 \\ &+ \frac{\kappa^2}{4}
\phi^2 \psi^2 \left( 1- \beta\frac{\psi^2}{2\Mpl^2}\right)^2.
\end{aligned}
\end{equation}
In the limit of negligible $\beta$, one recovers the
same potential as for the original hybrid model with $\lambda=
\lambda'=\kappa$, that is with a valley of local minima at
$\psi=0$. As $\beta$ increases, two symmetric valleys appear, parallel
to the central one as represented in Fig.~\ref{fig:shifted_potential_cut}.
These
new inflationary valleys get closer to the central one as $\beta$
gets larger. The central valley corresponds to a local
minima at large $\phi$. It becomes, as $\phi$ rolls toward
$0$, a local maxima at which point, two additional valleys appear.
Assuming a large $\phi$ field initially, inflation can be realized
along one of the three valleys. Several trajectories for various
initial conditions are represented on Fig.~\ref{fig:grid-shifted2}
to illustrate this. Inflation stops when the fields oscillate and
settle in one of the global minima of the potential.

\subsubsection{Space of initial conditions}
\begin{center}
\begin{figure}[ht]
\begin{center}
\scalebox{.45}{\includegraphics{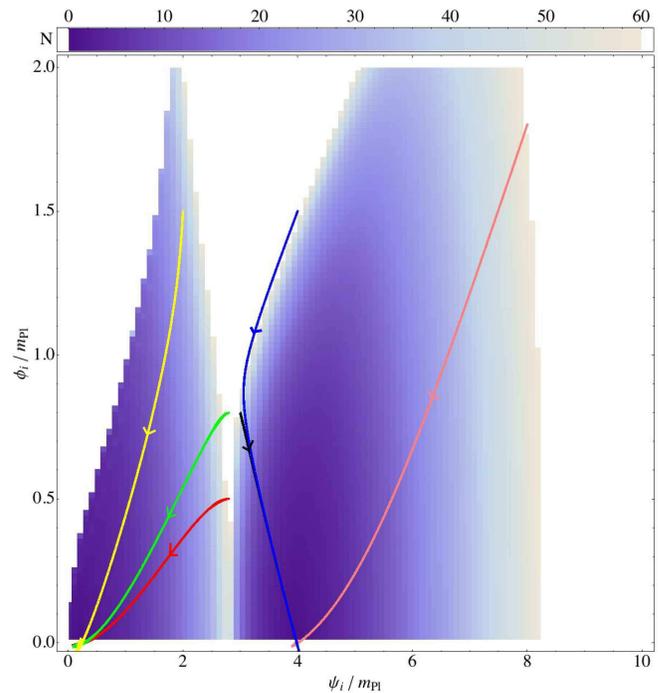}}
\caption{Grid of initial conditions leading or not to inflation,
for the shifted potential with $M=0.1 \mpl$, $\kappa=1$, and
$\beta = 10^{-2} \mpl^{-2}$. Some trajectories in field space have been
represented to identify where local maxima and minima are.} \label{fig:grid-shifted2}
\end{center}
\end{figure}
\end{center}

Grids of initial conditions leading or not to inflation have
been computed; one of them is represented in Fig.~\ref{fig:grid-shifted2}
for one set of parameters. It corresponds to one cut of the potential
in Fig.~\ref{fig:shifted_potential_cut} (dotted line).

For a small coupling $\beta$ (say of order $10^{-3}$), if we restrict
ourselves to values of the waterfall field smaller than $5\mpl$,
we obtain a space of initial conditions similar to the original
hybrid case (see Fig.~\ref{fig:completegrid-hybrid}), with a
triangular shaped region of unsuccessful inflation surrounded by
successful regions at higher values of the fields.

At larger values of $\psi$, around the new inflationary valley (the
``shifted'' one) at a positive $\psi$, a second triangular
shaped unsuccessful region is observed in addition.
For example, for $\beta=10^{-3}\mpl^2$, this shifted valley is located
at $\psi=9\mpl$ (at $\phi=2\mpl$). Unlike the
central one, the shifted valley is too steep to generate inflation when
the fields start inside it. Thus no line of successful initial
conditions along the valley is observed. Successful inflation is
only realized when starting sufficiently far from the valley, when the
potential becomes flatter around $\psi \in [3,8]\mpl$ (see
Fig.~\ref{fig:shifted_potential_cut}). \\

If we increase $\beta$, the shifted valley gets closer to
the $\psi=0$ one. As a consequence, the two unsuccessful regions
become closer as well, with interferences between them, as shown
in Fig.~\ref{fig:grid-shifted2}.
The shape of the first unsuccessful
region is modified because the presence of the second valley renders
some unsuccessful trajectories successful. We have represented some
examples of such trajectories in Fig.~\ref{fig:grid-shifted2}.
Finally, note that if this model is not considered as an effective
model for which fields can be super-planckian, in the limit of small
$\beta$ this model reduces to the original hybrid one. Conclusions
concerning the relative area of successful points are then the same.
For larger values of $\beta$ though, the new band of successful
inflation can appear even below the reduced Planck mass and increase
the probability of successful inflation. These results are
summarized in Tab.~\ref{tab:anamorph}.

\subsubsection{Supergravity corrections}
Let's discuss, as for the smooth hybrid model, the effects of
embedding the model in supergravity to study the robustness of our
conclusions under non-renormalizable corrections. As discussed in
the introduction, we remind the reader that neither supersymmetry
nor supergravity is a valid framework for describing
super-planckian fields and in this regime, the models studied are
considered as effective models. However supergravity corrections
allow to extend the domain of validity up to Planckian like field
values.

The supergravity corrections to the shifted potential are computed
assuming again a minimal K\"ahler potential and we obtain,
\begin{widetext}
\begin{equation}
\begin{split}
V_{\rm SUGRA}^{\rm sh}(S,\Phi,\bar{\Phi}) &= \kappa^2\mathrm{Exp}
\left[\frac{K_{\rm min}}{\Mpl^2}\right]\left\{ \left| \bar
\Phi \Phi -M^2  -\beta\frac{(\bar \Phi \Phi)^2 }{\Mpl^2} \right|^2\left(1-\frac{|S|^2}{\Mpl^2}
+\frac{|S|^4}{\Mpl^4}\right)\right.\\
&+|S|^2\left( |\Phi|^2 + |\bar \Phi|^2 \right)\left[\left|1-2\beta \frac{\bar
\Phi \Phi }{\Mpl^2}\right|^2+\frac{1}{\Mpl^4}\left| \bar
\Phi \Phi -M^2 -\beta\frac{(\bar \Phi \Phi)^2 }{\Mpl^2} \right|^2 \right]\\
&\left. + 2\frac{|S|^2}{\Mpl^2}\left[ \Phi \bar{\Phi}\left(1 - 2\beta\frac{\Phi
\bar{\Phi}}{\Mpl^2}\right) \left(\bar{\Phi}^* \Phi^*-M^2-\beta\frac{(\bar{\Phi}^* \Phi^*)^2}{\Mpl^2 }\right)+c.c.\right]\right\}~.
\end{split}
\end{equation}
\end{widetext}
By defining the inflaton and the waterfall field to be the
canonically normalized real part of the fields $S$, $\Phi$ and
$\bar{\Phi}$ like in the SUSY case, we obtain the effective 2-field
potential,
\begin{equation}\label{}
\begin{split}
&V_{\rm SUGRA}^{\rm sh} = \kappa^2 \mathrm{Exp}\left(\frac{\phi^2+\psi^2}{2\Mpl^2}\right)
\left\{\left(\frac{\psi^2}{4}-M^2-\beta \frac{\psi^4}{16\Mpl^2} \right)^2\right.\\
&\phantom{V_{\rm eff}^{\rm sh} = \kappa^2 \mathrm{Exp}^{\frac{\psi^2+\psi^2}{\Mpl^2}}
}~~~~~~~~~~~~\times\left(1-\frac{\phi^2}{2\Mpl^2}+\frac{\phi^4}{4\Mpl^4} \right)\\
&\left. + \frac{\phi^2 \psi^2}{4} \left[ 1- \beta \frac{\psi^2}{2 \Mpl^2}
+\frac{1}{\Mpl^2}\left(\frac{\psi^2}{4}-M^2-\beta \frac{\psi^4}{16\Mpl^2}\right)
\right]^2\right\}.
\end{split}
\end{equation}
These corrections affect at large initial values of the fields the dynamic
of inflation. At super-planckian initial values, the exponential term
dominates and the potential become too steep for inflation to be
automatically realized like in the SUSY case. However, the anamorphosis (or
type-C) trajectories still exist and are still the main origin of successful
inflation.
We have computed for several sets of the parameters the percentage
of successful initial conditions taking into account these corrections. We
don't find significant modifications compared to the SUSY case except at
large mass scale, where the steepness of the potential prevents from
inflation to be successful in the valleys. These results are summarized in
Tab.~\ref{tab:anamorph}.

\subsection{Radion Assisted Gauge Inflation}
\subsubsection{Motivations}
As mentioned in the introduction, the ``radion inflation''
model~\cite{Fairbairn:2003yx} belongs to the class of gauge inflation
models~\cite{ArkaniHamed:2003wu,ArkaniHamed:2003mz,Kaplan:2003aj}.
The main motivation of these models is to generate a sufficiently flat
inflaton potential protected by gauge symmetries because the inflaton
field is part of a gauge field.  As a consequence, it is safe to
consider super-planckian values for the inflaton field. Because its
potential is similar to the (smooth) hybrid one, this model is also
interesting to determine how generic the properties of initial
conditions observed for other models are, for different types of
models, originating from different high energy frameworks.

\subsubsection{The potential}
In the simplest version of these models, an effective 5-dimensional
universe is assumed, one of the dimension being compactified
with a radius\footnote{The effective 4-dimensional (reduced)
Planck mass is related to the 5-dimension Planck mass $M_5$ by
$\Mpl^2=2\pi R M_5^3$.} $R$. In the gauge inflation model, a gauge
symmetry is assumed together with a gauge field $(A_\mu,A_5)$. The
inflaton field is proportional to the phase $\theta$ of a
Wilson-loop wrapped around the compact dimension $\theta=\oint \dd x^5
A_5$. The full inflaton field is constructed with the symmetry
breaking scale $f$ of the gauge symmetry $\phi\equiv f\theta$. Its
potential is flat at tree level but at one-loop, takes the form of
an axion-like potential
\begin{equation}\label{eq:potengauge}
V(\phi)\propto \frac{1}{R^4}\cos (\phi/f)~.
\end{equation}

The potential is protected from non-renormalizable operators,
suppressed by powers of $1/R$, while non-perturbative quantum
gravity corrections can be
suppressed~\cite{ArkaniHamed:2003mz,Kaplan:2003aj}. Another
motivation concerns the initial homogeneity of the inflaton
field, necessary for inflation to start. Finally, since the
inflaton is a phase, one can show~\cite{Freese:1990rb} that the
probability to have a sufficiently homogeneous distribution of the
field is quite large.

The ``radion assisted'' gauge inflation differs from standard
gauge inflation by assuming a varying radius of the
extra-dimension $R$, around a central value $R_0$. The ``radion''
field is defined by\footnote{In our simulations below, we allowed
the $\psi$ field to take negative values because the symmetries of
the potential allow to redefine the field as $|\psi| \equiv (2\pi
R)^{-1}$, so that the length of the extra-dimension stays
positive.} $\psi \equiv (2\pi R)^{-1}$ and is subject to a
potential for which $R_0$ is assumed to be the minimum (for the
late time stability of the extra-dimension). The simplest way to
implement this stabilization is to use a Higgs-type potential for
$\psi$. By expanding, at first order, the potential of
Eq.~(\ref{eq:potengauge}), and by adding the Higgs-type sector,
the full scalar potential reads~\cite{Fairbairn:2003yx}
\begin{equation}\label{eq:potenradion}
V(\phi, \psi ) = \frac{1}{4}  \frac{\phi^2}{f^2} \psi^4 + \frac{\lambda}{4}
\left( \psi^2 - \psi_0 ^2 \right)^2~,
\end{equation}
where $\psi_0=(2\pi R_0)^{-1}$. This potential is similar to
the hybrid potential discussed in the last section. It is flat
for $\psi=0$ which corresponds to a global maxima. For a given
$\phi$, the minima of the potential are located in the valleys
\begin{equation}\label{ }
\langle\psi\rangle^2 = \frac{\psi_0^2}{1+\phi^2/(\lambda f^2)}~.
\end{equation}
More than $60$ e-folds of inflation can take place in these throats.


\subsubsection{Space of initial conditions}

Regarding the allowed parameter space that can be studied, since
$\psi$ is the inverse of the radius of an extra-dimension and
quantum gravity effects are expected to dominate when the field
gets larger than the five dimensional Planck mass. Thus
super-planckian values of $\psi$ or $\psi_0$ should not be taken
into account if one doesn't consider the potential of
Eq.~(\ref{eq:potenradion}) as an effective model. For the first set of
values of the parameters ($\psi_0 = 10^{-2}\mpl, f=\mpl,
\lambda=10^{-5}$), the grid of initial conditions is very similar
to the hybrid case, with a triangular unsuccessful region, and a
generic successful inflation at larger values of the fields (see
Fig.~\ref{fig:grilleRadion} below).
\begin{figure}[ht]
\scalebox{.45}{\includegraphics{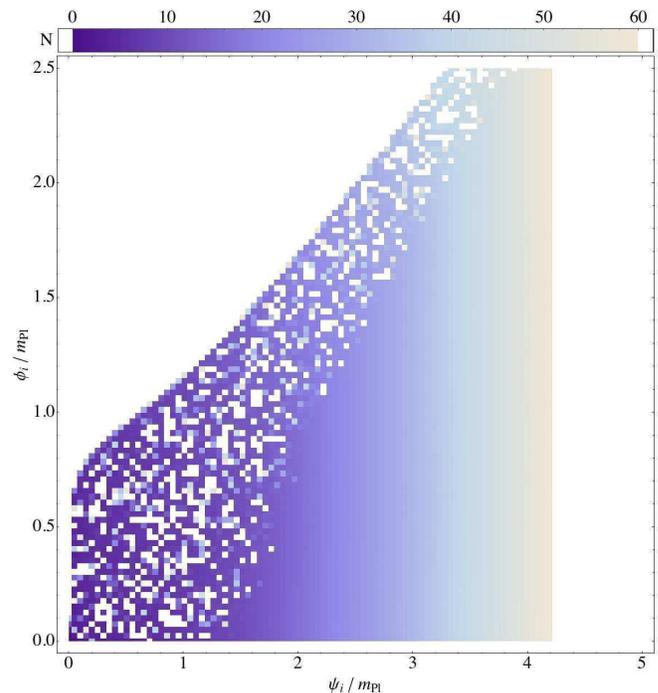}}
\caption{Grid of initial conditions for radion potential, with
$\psi_0=10^{-2} \mpl$, $f=1\mpl$, $\lambda=10^{-5}$.}
\label{fig:grilleRadion}
\end{figure}

Many successful trajectories also appear in the unsuccessful area
(type-C trajectories), for sufficiently small values of $\lambda
$. We observe a slightly higher successful area, compared to the
hybrid case: for $\phi_\ui,\psi_\ui< \Mpl$, more than $20\%$ of the
points are successful. Grids for different values of the parameter
$M$ show a behavior similar to the hybrid model. However, varying
$\lambda$ has a major impact on the amount of type-C trajectories
as shown in Tab.~\ref{tab:anamorph}. In particular we don't find a
significant amount of successful initial conditions for
for the choice of parameters of~\cite{Fairbairn:2003yx}
($\psi_0=10^{-2} \mpl$, $f=1 \mpl$, $\lambda = 10^{-3}$).
We also observe a transition between the successful and
unsuccessful region less abrupt (see Fig.~\ref{fig:grilleRadion}).
This is due to the fact that at small inflaton field, the potential
slightly differs from the hybrid potential: the slope of the
potential is slightly more steep and the same amount of e-folds
requires a larger variation of field values.

Our results on the proportion of successful initial conditions for
all models are summarized in the Tab.~\ref{tab:anamorph} below,
when restricting to initial fields values below the reduced Planck
mass. For comparison the results for the
original hybrid model are recalled. Two percentages are
given: first the total number of successful initial field values
(column 3) and the number of initial conditions that are scattered
in the initial condition space, outside of the inflationary valley(s)
(column 4). For these cases,
the realization of inflation is called in this paper ``anamorphosis'':
the system
fast-rolls down the potential, oscillates around the bottom of the
potential, climbs up one of the valleys and slow-rolls down along
it as if it started in the valley. When relevant, the number of the
figure representing the space of initial field values is given in
column 5.

\begin{table*}
\begin{center}
\begin{tabular}{|c|c|c|c|c|}
\hline Model & Values of parameters & Successful points (\%) &
Isolated points (\%) & Figure \\\hline\hline
Hybrid & $M=0.03 \mpl$, $m=10^{-6} \mpl$, $\lambda=\lambda'=1$ & 17  & 15 & \ref{fig:completegrid-hybrid} \\
Hybrid & $M = 0.06 \mpl$, $m=10^{-6}$, $\lambda=1$, $\lambda'=1$ & 11 & 6 &\\
Hybrid & $M=0.03 \mpl$, $m=10^{-5} \mpl$, $\lambda=\lambda'=1$ & 17 & 15 &\\
Hybrid & $M=0.03 \mpl$, $m=10^{-6} \mpl$, $\lambda=0.1$, $\lambda'=1$ & 16 & 14 & \\
Hybrid & $M=0.03 \mpl$, $m=10^{-6} \mpl$, $\lambda=1$, $\lambda'=0.1 $ & 3 & $ < 1$ &\ref{fig:hybridVaryLambda'}\\
Hybrid & $M = m =10^{-3} \mpl$, $\lambda=1$, $\lambda'=10^{-2}$ & 0 & 0 & \ref{fig:hybrid_redspectrum}\\\hline
Smooth & $M=10^{-2} \mpl$, $\kappa=1$  & 16 & 9 & \\
Smooth & $M=10^{-3} \mpl$, $\kappa=1$  & 53 & 49 & \\
Smooth & $M \approx 2.37\times 10^{-5}\mpl$, $\kappa\approx 10.3 $
& 78 & 60 & \ref{fig:smoothVaryAllLazarides} \\
Smooth SUGRA & $M=10^{-2} \mpl$, $\kappa=1$  & 29 & 17 & \\
Smooth SUGRA & $M=10^{-5}\mpl$, $\kappa=1 $ & 70 & 70 & \\\hline
Shifted & $M=0.1 \mpl$, $\kappa^2=1$, $\beta = 0.1 \mpl^{-2} $ & 6 & $ < 1$ &
\ref{fig:grid-shifted2}\\
Shifted & $M=10^{-2} \mpl$, $\kappa^2=1$, $\beta = 0.1 \mpl^{-2} $ & 15 & 14 &\\
Shifted & $M=10^{-2} \mpl$, $\kappa^2=1$, $\beta=1 \mpl^{-2} $ & 14 & 13 & \\
Shifted SUGRA & $M=0.1 \mpl$, $\kappa^2=1$, $\beta = 0.1 \mpl^{-2} $ & $ < 1$ & $ < 1$ &\\
Shifted SUGRA & $M=10^{-2} \mpl$, $\kappa^2=1$, $\beta = 0.1 \mpl^{-2} $ & 13 & 12 &\\
Shifted SUGRA & $M=10^{-2} \mpl$, $\kappa^2=1$, $\beta=1 \mpl^{-2} $ & 13 & 12 & \\ \hline
Radion & $\psi_0 = 10^{-2} \mpl$, $\lambda=10^{-3}$, $f=1\mpl $ & $<0.1$ & $<0.1$ & \\
Radion & $\psi_0 = 10^{-2} \mpl$, $\lambda=10^{-4}$, $f=1\mpl $ & 9.4 & 9.4 & \\
Radion & $\psi_0 = 10^{-2} \mpl$, $\lambda=10^{-5}$, $f=1\mpl $ &
25.6 & 24.8 & \ref{fig:grilleRadion}\\\hline
\end{tabular}
\caption{Percentage of successful points in grids of initial
conditions, for different models and values of parameters, when
restricting to $\phi_\ui,\psi_\ui\leq \Mpl$. The third column
represents the area of the whole successful initial condition
parameter space over the total surface. The fourth column
represents the surface of the successful space only located in
isolated points, over the total surface. Some of these sets are
represented in the body of the paper, the relevant figure being
reported in last column.} \label{tab:anamorph}
\end{center}
\end{table*}

From the different grids of initial values for the various models
studied in this paper, it is obvious that if we don't require that
the fields are smaller than the reduced Planck mass, the
proportion of successful initial conditions will tend toward
$100\%$ except when considering models in SUGRA. Therefore, we
have also conducted the same quantification
with the requirement $\phi_i,\psi_i <5 \mpl$. The results are given in
Tab.~\ref{tab:superplanck} below. This quantification has been
computed only to give an information about how fast the proportion
of successful initial conditions increases when the space of
allowed initial values is enlarged.
\begin{table}
\begin{center}
\begin{tabular}{|c|c|c|}
\hline Model & Values of parameters & Successful (\%)
\\\hline\hline
Hybrid &  $M=0.03,m=10^{-6},\lambda=\lambda'=1$ &   72 \\
Smooth &  $M \approx 2.37\times 10^{-5}\mpl$, $\kappa\approx  10.3 $ &   92 \\
Shifted & $M=10^{-2} \mpl $, $\kappa^2 = 1$, $\beta=10^{-2} \mpl^{-2}$  &  73 \\
Radion &  $\psi_0=10^{-2} \mpl$, $\lambda=10^{-3}$, $f=1$ & 76
\\\hline
\end{tabular}
\caption{Percentage of successful points in grids of initial
conditions of length $5\mpl$, for each model and some standard
values of the parameters.} \label{tab:superplanck}
\end{center}
\end{table}

\section{Conclusions}\label{sec:conclu}
Hybrid inflation is a class of models of inflation
motivated by high energy physics. In these models, the inflaton
field is assumed to be coupled to a Higgs-type auxiliary field
that ends inflation by instability, when developing a
non-vanishing expectation value. Two of its main
well-known problems - the blue spectrum of the non-supersymmetric
version of the model and the fine-tuning of the initial conditions
of the fields - are re-analyzed.

First, we found that the original hybrid model can generate a
red spectrum by two means.  As well-known, one way to have
inflation takes place in the large field phase is to have the
waterfall ending inflation in that phase. This requires a
constraint on the critical value of the inflaton triggering the
waterfall. We found a new criteria on the mass scale $\mu$ so
that a violation of the slow-roll conditions ensures the
\emph{non-existence of the small field phase of inflation}. In
both cases, the spectral index generated is less than unity (see
Fig.~\ref{fig:ns}). However, we show that this requires in both
cases a large initial value of the inflaton ($>\mpl$), and
therefore a realization of hybrid inflation in a regime away
from the limit $\phi \ll \mu$. This conclusion might reduce the
appeal of this model.

When considering the full two-field potential, it was
found~\cite{Tetradis:1997kp,Lazarides:1997vv,Mendes:2000sq} that
the original models suffer from a fine-tuning of the initial
values of the fields to generate a sufficiently long inflationary
phase. The space of successful inflation was thought to be
composed of a extremely narrow band along the inflationary valley
and some ``isolated scattered points'' which seemed randomly
distributed and of null measure~\cite{Mendes:2000sq}. It was
therefore considered that these models suffered from some
naturalness problem.

We have numerically integrated the exact equations of motion of
both fields and studied in details the space of initial
conditions. The study has been conducted for four different models
of hybrid-type inflation in various frameworks: the original
non-supersymmetric model (section~\ref{sec:originalhybrid}), its
extensions ``smooth'' and ``shifted'' hybrid inflation in global
supersymmetry and supergravity and the ``radion assisted'' gauge
inflation (section~\ref{sec:othermodels}). As expected, we found
that for sufficiently large initial values of the fields
(planckian-like or super-planckian), enough e-folds of inflation
are generated (see for e.g. Fig.~\ref{fig:completegrid-hybrid}).
This behavior is similar to the one of chaotic
inflation~\cite{Linde:1983gd}. We also studied the shape of the
unsuccessful region and its dependence on the potential
parameters: this property holds for any values of the parameters
and all the models we considered, except when embedded in
supergravity. Consequently, the first way to solve the fine-tuning
problem of initial conditions for hybrid models is to formulate
them in a framework for which it is safe to consider
planckian-like or super-planckian initial values for the fields.
This can be safe or problematic, depending on the framework used
to build the model as detailed in the introduction.

Even if the considered model is formulated in a framework where
the fields cannot safely be super-planckian, the unsuccessful
region of initial conditions contains successful sub-regions. They
correspond to special trajectories for which the velocity in field
space becomes oriented along the inflationary valley after some
oscillations at the bottom of the potential. Therefore the system
climbs up the valley before slow-rolling back down, generating
enough inflation. We find that these points form a complex
structure, as represented in Fig.~\ref{fig:anamorphosis}. They can
be seen as the anamorphosis of the standard inflationary valley,
and explain most of the successful initial conditions when
restricting to sub-planckian fields. The relative area that these
points occupy is typically of order of $15\%$ for the original
hybrid model. This value can go up to $25\%$ for radion inflation
and even above $70\%$ for smooth inflation, even though these
results depend on the values of the parameters of the potential
(see Tab.~\ref{tab:anamorph}). Moreover, even when supergravity
corrections are included, these trajectories still exist, their
proportion stays similar and they represent even more of the
successful initial conditions. We would like to note that these
percentages allow us to claim that the fine-tuning on hybrid
inflation is less severe that found in the past. However, to
translate this into an amount of fine-tuning for the model, it is
necessary to compute a measure of the probability space. This is
left for an extension of this work~\cite{Clesse:2009rr}.\\

Several other questions remain open and are extensions of this
work. We plan on investigating more deeply the statistical
properties of the anamorphosis regions of the plane of initial
conditions as well as the effects of initial velocities on this
plane~\cite{Clesse:2009r}. The study of the supersymmetric
versions of hybrid inflation, the F-term~\cite{Dvali:1994ms} and
D-term~\cite{Halyo:1996pp,Binetruy:1996xj} models are also left
for a future study. These models could have a different dynamics
from the models studied here since radiative corrections generate
potentially important corrections to the tree level
potential~\cite{Clauwens:2007wc}.

Finally, our results illustrated that the successful realizations
of hybrid inflation are not necessarily by fast-roll toward the
inflationary valley as usually assumed but also radially (type-B
trajectories), and in that sense chaotic like. Some aspects of the
phenomenology of these purely two-field trajectories (power
spectrum, generation of non-gaussianities, stochastic effects,
reheating, impact on topological defect formation) are still
unknown and should be studied.

\begin{acknowledgments}
It is a pleasure to thank J. Martin, C. Ringeval, and M. Tytgat
for many interesting discussions and a careful reading of the
manuscript. J. Garcia-Bellido and L. McAllister are also
acknowledged for useful comments. S.C. is supported by the Belgian
Fund for research (F.R.I.A.). J.R. is funded in part by IISN and
Belgian Science Policy IAP VI/11.
\end{acknowledgments}

\bibliographystyle{unsrt}
\bibliography{biblio}

\end{document}